\documentclass[referee]{raa}            

\usepackage{graphicx,times}             
\usepackage{natbib}
\usepackage{amssymb,amsmath}
\usepackage{inputenc}
\usepackage{array}

\bibpunct{(}{)}{;}{a}{}{,}

\usepackage[pagebackref=true]{hyperref}
\newcommand{\bjdtdb}{\ensuremath{\rm {BJD_{TDB}}}}

\newcommand\vsini{\ifmmode{v\sin{i_\star}}\else $v\sin{i_\star}$\fi}

\newcommand\sini{\ifmmode{\sin{i_\star}}\else $\sin{i_\star}$\fi}

\newcommand{\msun}{\ensuremath{M_{\rm \bigodot}}}
\newcommand{\rsun}{\ensuremath{R_{\rm \bigodot}}}
\newcommand{\lsun}{\ensuremath{L_{\rm \bigodot}}}
\newcommand{\mj}{\ensuremath{\,M_{\rm J}}}
\newcommand{\rj}{\ensuremath{\,R_{\rm J}}}

\usepackage{threeparttable}

\begin{document}

  \title{Confirmation of a Sub-Saturn-size transiting exoplanet orbiting a G dwarf: TOI-1194 b and a very low mass companion star: TOI-1251 B from TESS}

   \volnopage{Vol.0 (20xx) No.0, 000--000}      
   \setcounter{page}{1}          

   \author{Jia-Qi Wang 
      \inst{1,2}
   \and Xiao-Jun Jiang 
      \inst{1,2}
   \and Jie Zheng 
      \inst{1}
   \and Hanna Kellermann
      \inst{3,4}
   \and Arno Riffeser
      \inst{3}
   \and Liang Wang 
      \inst{2,5,6}
   \and Karen A. Collins
      \inst{7}
   \and Allyson Bieryla
      \inst{7}
   \and Lars A. Buchhave
      \inst{8}
   \and Steve B. Howell
      \inst{9}
   \and Elise Furlan
      \inst{10}
   \and Eric Girardin
      \inst{11}
   \and Joao Gregorio
      \inst{12}
   \and Eric Jensen
      \inst{13}
   \and Felipe Murgas
      \inst{14,15}
   \and Mesut Yilmaz
      \inst{16}
   \and Sam Quinn
      \inst{7}
   \and Xing Gao 
      \inst{17,18}
   \and Ruo-Yu Zhou 
      \inst{19}
   \and Frank Grupp
      \inst{3,4}
   \and Hui-Juan Wang 
      \inst{1,2}
   }
   \institute{
            CAS Key Laboratory of Optical Astronomy, National Astronomical Observatories, Chinese Academy of Sciences, Beijing 100101, China \url{wanghj@nao.cas.cn}\\
        \and
            University of Chinese Academy of Sciences, 19A Yuquan Road, Shijingshan District, Beijing 100049, China\\
        \and
            University Observatory Munich, Scheinerstrasse 1, 81679 Munich, Germany\\
        \and
            Max Planck Institute for Extraterrestrial Physics, Giessenbachstr. 1, 85748 Garching, Germany \url{frank@grupp-astro.de}\\
        \and
            National Astronomical Observatories/Nanjing Institute of Astronomical Optics $\&$ Technology, Chinese Academy of Sciences, Nanjing 210042, China\\
        \and
            CAS Key Laboratory of Astronomical Optics $\&$ Technology, Nanjing Institute of Astronomical Optics $\&$ Technology, Nanjing 210042, China
        \and
            Center for Astrophysics $\mid$ Harvard $\&$ Smithsonian, 60 Garden Street, Cambridge, MA 02138, USA\\
        \and
            National Space Institute, Technical University of Denmark, Elektrovej, 2800 Kgs. Lyngby, Denmark\\
        \and
            NASA Ames Research Center, Moffett Field, CA 94035, USA\\
        \and
            NASA Exoplanet Science Institute, Caltech/IPAC, Mail Code 100-22, 1200 E. California Blvd., Pasadena, CA 91125, USA\\
        \and
            Grand Pra Observatory, Switzerland\\
        \and
            Department of Physics, Washington University, St. Louis, Missouri 63130, USA\\
        \and
            Department of Physics and Astronomy, Swarthmore College, Swarthmore, Pennsylvania 19081, USA\\
        \and
            Instituto de Astrofísica de Canarias (IAC), 38205 La Laguna, Tenerife, Spain\\
        \and
            Departamento de Astrofísica, Universidad de La Laguna, Tenerife, Spain\\
        \and
            Dept. of Astronomy and Space Sciences, Ankara University, Tandogan 06100 Ankara, Turkey\\
        \and
            Xinjiang Astronomical Observatory, Chinese Academy of Sciences, Urumqi 830011, China\\
        \and
            Urumqi No.1 Senior High School, Urumqi 830002, China\\
        \and
            Lijiang Gemini Observatory, Lijiang 674199, China\\
\vs\no
   {\small Received 20xx month day; accepted 20xx month day}
}

\abstract{We report the confirmation of a sub-Saturn-size exoplanet, TOI-1194 b with a mass about $0.456_{-0.051}^{+0.055}$ $M_{J}$, and a very low mass companion star with a mass of about $96.5\pm1.5$ $M_J$, TOI-1251 B. Exoplanet candidates provided by the Transiting Exoplanet Survey Satellite (TESS) are suitable for further follow-up observations by ground-based telescopes with small and medium apertures. The analysis is performed based on data from several telescopes worldwide, including telescopes in the Sino-German multiband photometric campaign, which aimed at confirming TESS Objects of Interest (TOIs) using ground-based small-aperture and medium-aperture telescopes, especially for long-period targets. TOI-1194 b is confirmed based on the consistent periodic transits depths from the multiband photometric data. We measure an orbital period of $2.310644\pm0.000001$ d, and radius is $0.767_{-0.041}^{+0.045}$ $R_J$, and amplitude of RV curve is $69.4_{-7.3}^{+7.9}$ m/s. TOI-1251 B is confirmed based on the multiband photometric and high-resolution spectroscopic data, whose orbiting period is $5.963054_{-0.000001}^{+0.000002}$ d, the radius is $0.947_{-0.033}^{+0.035}$ $R_J$, and amplitude of RV curve is $9849_{-40}^{+42}$ m/s.
}

   \authorrunning{Wang et al.}            
   \titlerunning{TOI-1194 b and TOI-1251 B}  
   \maketitle

\section{Introduction} \label{sec:intro}

Since the discovery of PSR1257 + 12 c and d in 1992 \citep{1992Natur.355..325R}, more than 5,000 planets have been confirmed. Ground-based and space-based telescopes have significantly contributed to exoplanet detection and research. The expected efficiency of the transit method for exoplanet search is very optimistic\citep{2003ASPC..294..361H}, but the actual situation in recent years shows that the rate of planet identification and the physical characteristics of the discovered planets are obviously limited by the detection methods and observation strategies. For instance, about 17\% confirmed planets revolve around their host star with an orbital period greater than 100 days\footnote{According to NASA Exoplanet Archive\citep{2013PASP..125..989A}, \url{https://exoplanetarchive.ipac.caltech.edu/docs/counts_detail.html}}, which makes it more challenging to detect long-period and wide-orbit planets. These planets share similarities with our own solar system and require further research. Understanding these exoplanets and their properties can yield valuable insights into planetary formation and evolution. Therefore, the immediate goal is to expand the sample of detected planets through advanced detection methods and observation strategies.

The most efficient way to confirm exoplanets is through space-based surveys and follow-up studies from ground-based telescopes. The nine-year survey by the Kepler mission \citep{2010Sci...327..977B, 2014PASP..126..398H} gives us a glimpse into the abundance of planets in the universe. Since the launch of TESS \citep{2015JATIS...1a4003R} in Apr. 2018, numerous exoplanet candidates have been discovered and confirmed. However, for some candidates with extended periods ($\geq$ 27 days), or only a single transit detected (called single-transit \citep{2018A&A...619A.175C}) require further confirmation and research, which may demand a considerable amount of observation time at a single ground-based site. In addition, capturing the full transit can be challenging in certain cases. A well-designed multi-telescope campaign provide a solution to this problem, and the feasibility of this approach has been validated by the studies on long-period planet HD 80606 b, which has a period of 111.43605 days\citep{2001A&A...375L..27N,2009A&A...502..695P}. For instance, telescopes located at 40 degrees latitude typically have an average dark time of around 8 hours per observation night. In this case, multiple telescopes spanning a time zone of fewer than eight hours can work together to perform relay observations on exoplanets with transit durations exceeding eight hours. By doing so,  more data can be obtained with a single transit for these planets, which have larger orbital radii and a higher probability of being in the habitable zone.

Simultaneous multiband photometry using ground-based telescopes provides a cost-effective and efficient method for identifying exoplanet candidates and reducing false positives (FPs). While space-based telescopes capture the decline in candidate stars' luminosity through long-term sky area monitoring, the satellite orbit and observation conditions limit the available observation time to a single band, ensuring survey efficiency and the robustness of the telescope system. It is worth noting that over half of the exoplanet candidates are subject to stellar luminosity variations or are eclipsing binaries (EBs) \citep{2018haex.bookE..79D}. Simultaneous multi-color light curves can effectively eliminate most FPs caused by EBs\citep{2004A&A...425.1125T}. In contrast, the Kepler mission has shown that low transit signal-to-noise ratios (SNRs) are more likely to result in FPs, particularly in the case of Earth-like planets \citep{2013ApJ...766...81F}. Although this issue also affects ground-based telescope detection, the ample observation time provides an opportunity to accumulate more time-series data of transits and enhance the transit SNRs by adding multiple transits \citep{2009ApJ...698..519K}.

Since the first transit planet was confirmed\citep{2000ApJ...529L..45C}, photometry of transits has become the most efficient method for identifying new planet candidates and determining their radius ($R_P$) and semi-major axis (a) \citep{2010exop.book...55W}. Achieving 1\% precision in radius determination is essential for studies of planetary structure and evolution, and relies heavily on accurate measurements of stellar limb darkening values \citep{2013A&A...549A...9C}. Transit light curves morphology, particularly the ingress and egress fragments, can significantly impact model-estimated parameters. The transit observation cadence is the primary driver to such morphology \citep{2010MNRAS.408.1758K}. Ground-based telescopes with meter-scale apertures (called as one meter-class telescope) enable high SNRs$\geq$100 for follow-up observations, while maintaining an observation cadence that conforms to exposure time constraints.

We want to present some of our latest confirmations of TOI-1194 b and TOI-1251 B through our global follow-up campaign utilizing several telescopes, including the Sino-German multiband photometric campaign. Section \ref{sec:campaign} offers an introduction to the Sino-German multiband photometric campaign, target selection and the follow-up observation strategy employed during our campaign. Section \ref{sec:Reduction} described the rapid processing of intensive data generated by campaign telescopes and light curve fitting, while Section \ref{sec:Verify} details our confirmation of TOI-1194 b and TOI-1251 B. Lastly, Section \ref{sec:sum} concludes with a summary and discussion of our findings.

\section{Sino-German Multiband Photometric Campaign} \label{sec:campaign}
\begin{figure}[htb]
\centering
\includegraphics[width=16.2cm, height=9cm]{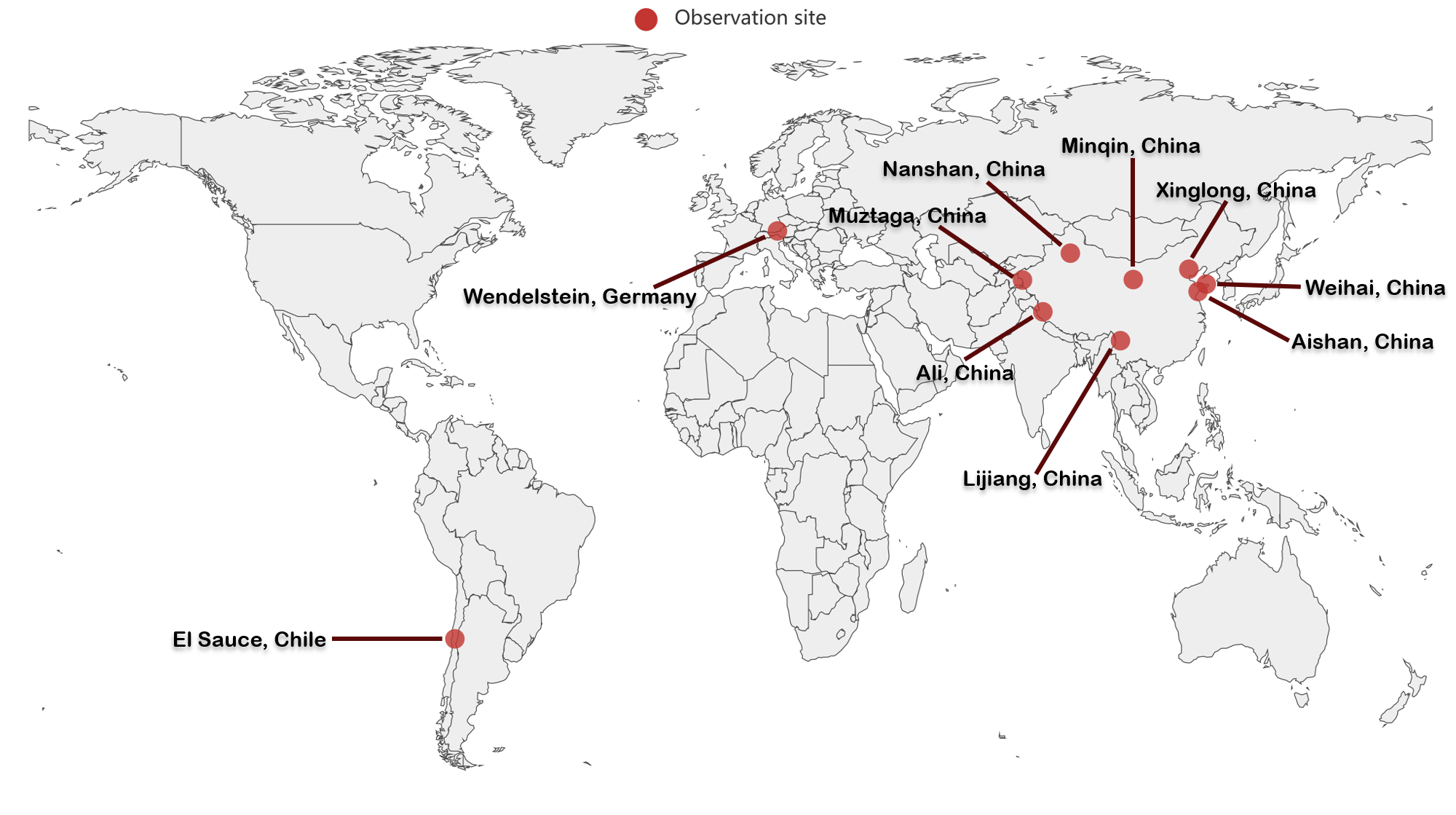}
\caption{Location of each site in the campaign}
 \label{sites}
\end{figure}

In order to identify a greater number of long-period planets, our Sino-German cooperation team has successfully established a global multiband follow-up observation network. This network incorporates a wide range of optical observational equipment, including both professional-grade instruments and those utilized by amateur astronomers. The network remains distributed across several telescopes in China, Germany, and Chile, covering a time zone spanning five to twelve hours among the sites, as shown in Figure \ref{sites}. Currently, the network includes fifteen telescopes located at ten observation sites.

Our follow-up campaign utilizes multiband photometric observations to exclude FPs from TOIs we have observed. Nonetheless, filter systems may differ among several telescopes. For instance, telescopes at Xinglong \citep{2016PASP..128k5005F,2018RAA....18..107B, 2012RAA....12.1585H} and Weihai \citep{2016PASP..128l5002G, 2014RAA....14..719H} in China use the Johnson-Cousins filter system, while telescopes at Weldelstein Observatory in Germany \citep{2014SPIE.9145E..2DH, 2021SPIE11823E..1US} and Nanshan in China use the \textup{SDSS} $u'g'r'i'z'$ system. Other amateur telescopes use RGB filters or white band (refer to Table \ref{tab:site}).

\begin{table}[htb]
\centering
\caption[]{Primary Parameters of Telescopes in Sino-German Campaign \label{tab:site}}
\begin{tabular}{c c c c c c c}
 \hline
 Site & Longitude & Latitude & Altitude & Aperture & filter & FoV. \\
  & ($^\circ$) & ($^\circ$) &  (m) & (cm) &  &  \\
 \hline
 Nanshan, China & 87.1736E & 43.4747N & 2088 & 25/60 & \textup{SDSS } ${g'r'i'}$ & 62\arcmin/25\arcmin \\
 Aishan, China & 119.9453E & 36.1231N & 100 & 10/25 & \textup{LRGB} &  50\arcmin/87\arcmin \\
 Weihai, China & 122.0496E & 37.5359N & 100 & 60/100 & \textup{Johnson BVRI} & 30\arcmin/12\arcmin \\
 Lijiang, China & 100.0300E & 26.6951N & 3200 & 30/35 & \textup{SDSS } ${g'r'i'}$ /\textup{LRGB} & 21\arcmin/25\arcmin \\
 El Sauce, Chile & 70.7631W & 30.4725S & 1525 & 35/60 & \textup{LRGB} &  60\arcmin/60\arcmin \\
 Xinglong, China & 117.5772E & 40.3958N & 950 & 60/80/85 & \textup{Johnson BVRI} &  37\arcmin/11\arcmin/30\arcmin \\
 Wendelstein, Germany & 12.0121E & 47.7364N & 1838 & 43 & \textup{SDSS} ${g'r'i'}$ & 43\arcmin\\
 Ali, China & 80.1272E & 32.2312N & 5360 & 50 &\textup{Johnson BVRI}& 11\arcmin \\
 Muztaga, China & 74.8967E & 38.3297N & 4583 & 50 &\textup{Johnson V} & 5\arcmin \\
 Minqin, China & 103.3152E & 38.4141N & 1335 & 60 &\textup{Johnson BVRI} & 30\arcmin \\
 \hline
\end{tabular}
\end{table}

\subsection{Targets Selection and Observation Strategy} \label{sec:Strategy}

To confirm exoplanets, obtaining high-quality data depends upon choosing candidates with full-transit when observation conditions are favorable. More than half of the TOI targets have brightness above 12 magnitude in the $TESS$ band, making them suitable for observation using one meter-aperture telescopes. During the Sino-German campaign follow-up observation, our strategy aims to maximize the observation efficiency within the limited observation time. Our team achieves this efficient observation strategy by calculating all TOIs transits for each observation site during the observable time. The TOIs are then ranked priorities based on observation parameters such as brightness, transit depth, and orbital period. Between Oct. 2019 to May. 2023, we screened and observed a total of 181 TOIs. Through our follow-up data, we confirmed a planet orbiting a solar-type star, TOI-1194 b, and a very low-mass stellar companion, TOI-1251 B.

For one meter-aperture telescopes, the main factor affecting the precision of photometric observation data is SNR. Point source standard SNR formula is as follows:

\begin{align*}\label{SNR_ground}
SNR = \frac{R_{*} \times t} {\sqrt{R_{*} \times t + n_{pix} \times (R_{sky} \times t + {RN}^{2} + D \times t)}}
\end{align*}

In this context, $R_{*}$ denotes the target flux, $t$ represents exposure time, $R_{sky}$ indicates the sky background flux, $n_{pix}$ signifies the number of pixels occupied by star, $RN$ refers to camera readout noise and $D$ denotes dark current \citep{2006hca..book.....H}. Due to the use of camera cooler with temperature below -80 $^\circ C$, dark current noise can be ignored. A faster readout speed results in greater readout noise, while a slower readout speed tends to be preferred. Analyses of Corot and Kepler data by Kipping have revealed that sampling frequency has a particularly crucial effect on planets confirmation and the determination of planetary parameters\citep{2010MNRAS.408.1758K}. Longer exposure cadences of photometry images can result in increasing levels of systematic errors concerning physical parameters. On the 60cm telescope at the Xinglong Observatory, we employed the ANDOR \#DZ936N CCD camera with an SNR of 100 for a 12-mag star exposure lasting 10 seconds. By utilizing a readout speed of 1MHz, the readout time per image was reduced to 4 seconds, albeit with a readout noise of 6.5 electrons. Conversely, the readout time was prolonged to 76.8 seconds and the readout noise reduced to 3.6 electrons at a readout speed of 50kHz. Furthermore, we selected a gain value closest to 1 to curb readout noise linked to amplification. The CCD/CMOS exposure parameters were tailored to meet these standards and set for the telescope used during our campaign. Consequently, the photometric data obtained achieved an accuracy better than 0.002 magnitudes when the SNR was no less than 100.

It is noteworthy that roughly 20\% of the TOIs brighter than eight in $TESS$ band, making them well-suited for observation with one meter-class telescopes. However, when observing TOIs brighter than 8 mag, the FoV of one-meter telescopes is typically limited to approximately 30\arcmin, making identification of reference stars with similar brightness within the FoV difficult. To ensure optimal data quality, it is imperative to appropriately defocus the target image in order to avoid saturation, while also limiting the exposure time to effectively enhance the SNR of fainter reference stars within the FoV. The efficacy of observation strategy is critical to successful data acquisition.

\section{Observations} \label{sec:Obs}

We observed two planet candidates from TESS, TOI-1194.01 and TOI-1251.01 by using several ground-based one meter-class telescopes in Sino-German campaign and TESS Follow-up Observing Program (TFOP)\footnote{\url{https://tess.mit.edu/followup/}}. The photometric observation parameters were summarized in Table \ref{tab:TOI1194_obs} and \ref{tab:TOI1251_obs}, followed by a description of the observations by different observatory sites and TESS.

\subsection{TESS Observation}
Sector 14 of TESS observation marked the first time discovery of TOI-1194 b and TOI-1251 B. TOI-1194 b was alerted as a candidate on Aug. $27^{\rm{th}}$, 2019, with 2-min cadence data  subjected to reduction by Science Processing Operations Center (SPOC) pipeline \citep{2016SPIE.9913E..3EJ}. The resulting output showed evidence of a companion with a period of 2.31 days. We performed a check of the Simple Aperture Photometry (SAP) fluxes \citep{2010SPIE.7740E..23T,2020ksci.rept....6M} for TOI-1194 b in Sector 14, 20, 21, 40, 41, 47, 48 and found no trace of additional companions.

TOI-1251 B was flagged as a candidate on Oct. $17^{\rm{th}}$, 2019, after reducing its 2-min cadence data with the SPOC pipeline. Its high ecliptic latitude resulted in visibility across Sector 14-26. The light curves indicated a companion with a period of 5.96 days. We searched for additional companions using the SAP fluxes in Sector 14-26 and 40-41,48-49 and 51-52, but no others were detected.

\subsection{Ground-Based Follow-up Photometric Observation} \label{sec:followup}

\subsubsection{Xinglong Observatory}
We obtain data from three telescopes at the Xinglong Observatory, Hebei, China, during the Sino-German follow-up campaign. Specifically, data from the 0.6m telescope, TNT (0.8m) \citep{2012RAA....12.1585H}, and NBT (0.85m) \citep{2018RAA....18..107B} resulted in the acquisition of eight series of data (seven primary transits and one secondary transit) of TOI-1194 b and two series of data (all primary transit) for TOI-1251 B. We require that the SNRs of the images are greater than 1000 pixel$^{-1}$, and the exposure times of each band were adjusted according to environmental conditions such as sky background and seeing, which are not a fixed value.

The selected CCD exposure parameters are outlined in Table \ref{tab:CCD_par}. Despite our efforts, the primary transit on Dec. $2^{\rm{nd}}$, 2021 and secondary transit on Mar. $1^{\rm{st}}$, 2022 of TOI-1194 b could not be satisfactorily captured due to poor photometric accuracy. Analysis of planetary parameters and properties was primarily based on these transits data outlined in Table \ref{tab:TOI1194_obs} and \ref{tab:TOI1251_obs}, which details the observation date, filters, and the number of images.

\begin{table}[htb]
\centering
\caption[]{CCD Parameters of Telescopes at Xinglong Observatory Used in TOIs Follow-Up Observations \label{tab:CCD_par}}
\begin{tabular}{c c c c c}
  \hline
  Telescope & CCD type & Readout noise & Gain & Pixel scale \\
   & & ($e^{-}$ rms) & ($e^{-}$ per A/D) & (\arcsec pixel$^{-1}$)\\
  \hline
  60cm & Andor DZ936N & 6.5 & 1.1 & 1.10\\
  80cm & PYL1300BX & 4.02 & 1.36 & 0.51\\
  85cm & Andor DZ936N & 7.0  & 0.97 & 0.93\\
  \hline
\end{tabular}
\end{table}

\subsubsection{Canela's Robotic Observatory (CROW)}
We conducted an observation of TOI-1194 b on Dec. $28^{\rm{th}}$, 2019 in the \textup{SDSS} $g^\prime$ band using the CROW 0.35m telescope located in Portugal. The telescope is equipped with an ST-10XME camera and a pixel scale of 0.66\arcsec. The exposure time of each image is 180s and SNR value of target star greater than 850 pixel$^{-1}$. We obtained 58 data points throughout the entire transit, which yielded photometric data analyzed using \texttt{AstroImageJ} \citep{2017AJ....153...77C} with rms = 0.001 mag precision.

\subsubsection{Grand-Pra Observatory (GdP)}
Our observation of TOI-1194 b on Dec. $5^{\rm{th}}$, 2019, was conducted in the \textup{SDSS} $z^\prime$ band using the RCO 0.4m telescope located at the GdP in Switzerland. The telescope feathers a FLI4710 camera with a pixel scale of  0.73\arcsec. The exposure time is 150s and SNR value of target star greater than 540 pixel$^{-1}$. We acquired 107 data points throughout the entire transit, enabling successful photometric analysis using \texttt{AstroImageJ} \citep{2017AJ....153...77C} with rms = 0.002 mag precision.

\subsubsection{Fred Lawrence Whipple Observatory (FLWO)}
On Jan. $14^{\rm{th}}$, 2020, we observed TOI-1194 b in the B band using the KeplerCam instrument mounted on the FLWO 1.2m telescope, located at Mt. Hopkins in Arizona, USA. KeplerCam is a wide-field CCD camera constructed to provide multiband follow-up photometry for the Kepler Input Catalog (KIC), with a pixel scale of 0.67\arcsec; The four output amplifiers are read out in unison at 200 kHz;  the total readout time and instrument overhead are 11 seconds when binned 2$\times$2 \citep{2005AAS...20711010S}. The exposure time is 15s and SNR value of target star greater than 300 pixel$^{-1}$. We acquired 424 data points throughout the entire transit, which enabled us to conduct successful photometric analysis using \texttt{AstroImageJ} \citep{2017AJ....153...77C} with rms = 0.002 mag precision.

\subsubsection{Peter Van De Kamp Observatory (PvdK)}
We observed TOI-1194 b on Oct. $5^{\rm{th}}$, 2019, in the \textup{SDSS} $i^\prime$ band by using PvdK telescope, an f/7.8 Ritchey-Chretien telescope located in Pennsylvania, USA, with a aperture of 0.6m. The telescope employed an Apogee U16M camera, and the pixel scale is 0.76\arcsec with 2$\times$2 binning \citep{2012ApJ...756L..39B}. The exposure time is 98s and SNR value of target star greater than 350 pixel$^{-1}$. The observation spanned the entire transit and yielded 58 data points, which were processed using \texttt{AstroImageJ} \citep{2017AJ....153...77C} with rms = 0.002 mag precision.

\subsubsection{Ankara University Kreiken Observatory (AUKR)}
We observed TOI-1251 B using the 0.8m Prof. Dr. Berahitdin Albayrak Telescope (hereafter T80 Telescope) on Oct. $12^{\rm{th}}$, 2019 and on Aug. $17^{\rm{th}}$, 2020 from the AUKR Observatory in Turkey, both observations encompassed the entire transit. The T80 telescope is equipped with an APOGEE ALTA U47 camera, the pixel scale is 0.8\arcsec. The filter used on Oct. $12^{\rm{th}}$, 2019 is R band, and the exposure time is 12s that SNR value of target star greater than 200 pixel$^{-1}$. The observation on Aug. $17^{\rm{th}}$, 2020 used \textup{SDSS} $g^\prime$ and $z^\prime$ bands, the exposure time in $g^\prime$ is 10s that SNR value of target star greater than 100 pixel$^{-1}$ and the exposure time in $z^\prime$  is 30s that SNR value of target star greater than 250 pixel$^{-1}$. The images were analyzed using \texttt{AstroImageJ} \citep{2017AJ....153...77C} with rms = 0.004 mag precision in each band.

\subsubsection{Teide Observatory (OT)}
We observed TOI-1251 B on Oct. $18^{\rm{th}}$, 2019 in the \textup{SDSS} $g^{\prime}r^{\prime}i^{\prime}z^{\prime}$ bands using the MuSCAT2 camera mounted on the 1.52 m Telescopio Carlos S\'anchez (TCS) telescope located at the OT in Tenerife, Spain. The primary mirror is fixed in an equatorial structure with a Cassegrain focus and focal length of f/13.8 in a Dall-Kirkham type configuration. MuSCAT2 has a capability of 4-color simultaneous imaging in \textup{SDSS} $g^{\prime}r^{\prime}i^{\prime}z^{\prime}$, with a pixel scale of 0.44\arcsec \citep{2019JATIS...5a5001N}. The exposure time is 3s in each band that SNR value great than 300 pixel$^{-1}$. Our observation covers full transit and yielded 765 points in each band, which were analyzed using \texttt{AstroImageJ} \citep{2017AJ....153...77C} with the best rms = 0.004 mag precision.

\begin{table}[htb]
\centering
\caption[]{Ground-Based Photometric Follow-Up Observation of TOI-1194 b \label{tab:TOI1194_obs}}
\begin{threeparttable}
\begin{tabular}{c c c c c}
  \hline
  Date & Telescope & Bands & Duration & Number of images\\
   & & & (min) & \\
  \hline
  2019/10/16 & PvdK 0.6m & SDSS $i^\prime$  & 200 & 58 \\
  2019/12/05 & GdP 0.4m & SDSS $z^\prime$  & 295 & 107 \\
  2019/12/28 & CROW 0.35m & SDSS $g^\prime$  & 249 & 58 \\
  2020/01/14 & FLWO 1.2m & B  & 244 & 424 \\
  2020/04/22 & Xinglong 0.8m & BVRI  & 246 & 314 \\
  2021/03/23 & Xinglong 0.6m & BVRI  & 149 & 182 \\
  2021/04/06 & Xinglong 0.6m & BVRI  & 128 & 190 \\
  2021/04/13 & Xinglong 0.8m & BVRI  & 144 & 424 \\
  2021/12/02$^{1}$ & Xinglong 0.8m & BVRI  & 75 & 100 \\
  2022/02/21 & Xinglong 0.85cm & BVRI  & 131 & 720 \\
  2022/02/28 & Xinglong 0.6m & BVRI  & 138 & 156 \\
  2022/03/01$^{2}$ & Xinglong 0.6m & BVRI  & 165 & 476 \\
    &  &  &  & Total images: 3209 \\
  \hline
\end{tabular}
  \begin{tablenotes}
    \footnotesize
    \item[*] Ignored because of poor photometric quality.
    \item[**] Secondary transit.
  \end{tablenotes}
\end{threeparttable}
\end{table}

\begin{table}[htb]
\centering
\caption[]{Ground-Based Photometric Follow-Up Observation of TOI-1251 B \label{tab:TOI1251_obs}}
\begin{tabular}{c c c c c}
  \hline
  Date & Telescope & Bands & Duration & Number of images\\
   & & & (min) & \\
  \hline
  2019/10/12 & AUKR 0.8m & R & 315 & 1267 \\
  2019/10/18 & TCS 1.52m & SDSS $g^{\prime}r^{\prime}i^{\prime}z^{\prime}$ & 77.2 & 3060 \\
  2020/04/02 & Xinglong 0.6m & BVRI  & 257 & 807 \\
  2020/08/17 & AUKR 0.8m & SDSS $g^\prime$,$z^\prime$  & 363 & 706 \\
  2021/08/28 & Xinglong 0.85m & BVRI  & 243 & 845 \\
   &  &  &  & Total images: 6685 \\
  \hline
\end{tabular}
\end{table}

\subsection{Ground-Based Follow-up Spectroscopic Observation}

\subsubsection{Xinglong Observatory}
From 2021 to 2022, we acquired 14 high-resolution spectra of TOI-1251 B on five separate nights by using the fiber-fed High Resolution Spectrograph(HRS) \citep{2016PASP..128k5005F} mounted on Xinglong 2.16m telescope. The HRS is a high-resolution echelle spectrograph, equipped with an astro-comb \citep{2019MNRAS.482.1406Z} for high precision wavelength calibration, a resolution R=49,000 and a spectral range spanning 360-1000nm. While ensuring time resolution, we aimed to optimize the exposure time to acquire data with high SNRs. Therefore, a single exposure not exceeding 1800s for targets brighter than V = 11.5. We used spectra in 600-700nm to determine RVs, with exposure time of 1800s we yielded SNRs in 28-33 pixel$^{-1}$ in this wavelength range. All spectra were exposed with simultaneous calibration fiber that calibration spectra were simultaneously obtained with the target in one image to reduce any potential casual error between different images.

\subsubsection{Roque De Los Muchachos Observatory (ORM)}
We conducted observations of TOI-1251 B using the high-resolution FIbre-fed Echelle Spectrograph (FIES) \citep{2014AN....335...41T} instrument mounted on Nordic Optical Telescope (NOT, 2.56m) located at ORM, La Palma, Spain \citep{2010ASSP...14..211D}. FIES is a cross-dispersed high-resolution echelle spectrograph that employs two fibers simultaneously feeding the spectrograph. The entire spectral range covers 376-884nm, with our spectra obtained in HIGH-RES mode where R=67,000. We used the wavelength range from approximately 400–550nm to determine the RVs. The exposure time was approximately 1200s yielding an SNR in 30-40 pixel$^{-1}$ in the wavelength range used. From Jun. $26^{\rm{th}}$ to Oct. $11^{\rm{th}}$, 2020, we obtained 12 spectra of TOI-1251 B totally.

\subsubsection{FLWO}
We utilized the Tillinghast Reflector Echelle Spectrograph (TRES) \citep{Frsz2008DESIGNAA} mounted on 1.5 m Tillinghast Reflector telescope at FLWO to observed both TOI-1194 b and TOI-1251 B. TRES visible range is covers 385-909nm at a resolution of R=44,000. From Dec. $12^{\rm{th}}$-$13^{\rm{th}}$, 2019, and Feb. $5^{\rm{th}}$-$26^{\rm{th}}$, 2023, we acquired twelve spectra of TOI-1194 b with exposure times ranging from 650s to 2000s, and SNR approximate to 30 pixel$^{-1}$. For TOI-1251 B, we obtained two spectra with exposure times of 1500s (SNR$\sim$40 pixel$^{-1}$) and 900s (SNR$\sim$33 pixel$^{-1}$), respectively, on Oct. $20^{\rm{th}}$ and $23^{\rm{rd}}$, 2019.

\section{Photometric Data Analysis} \label{sec:Reduction}

\subsection{Data Reduction Pipeline: QLCP} \label{sec:pipeline}
Following the acquisition of observation follow-up data, we performed data reduction and utilized it to identify the transit signals. To increase the sampling frequency of key transit phase, such as ingress and egress, we shortened the exposure time of each image while ensuring a high SNR. We employed our Quick Light-Curve Pipeline (QLCP) \citep{2023ART...01...83}, a Python-based pipeline, QLCP, has the capability of performing differential photometry and generating light curves by aligning stars from a series of photometric images, along with conducting automated bias and flat corrections. Fluxes are calculated by default using \texttt{MAG\_AUTO} of Source-Extractor \citep{1996A&AS..117..393B}, with the option of switching to alternative flux evaluation methods by adjusting the configuration. Field stars with stable fluxes be identified by QLCP as reference stars, or they can be manually designated. The pipeline generates a list in both text and binary formats, along with locating charts and producing light curve figures for further analysis.

\subsection{Creation of Light Curve} \label{sec:lc}
For follow-up photometric images of TOIs with $TESS$ magnitudes between nine to twelve, at least five reference stars can be selected within a 30\arcmin field of view (FoV), provided that the brightness difference between the reference stars and the target does not exceed two magnitudes. QLCP is capable of conducting high-precision differential photometry for reference stars with SNRs close to 100 despite significant difference in brightness, by utilizing adaptive aperture photometry in Sextractor \citep{1996A&AS..117..393B}. Therefore, it's very suitable for processing the transit data of TOIs brighter than 8 magnitude which were observed according to the strategy described in Section\ref{sec:Strategy}. We use QLCP to quickly generate light curves after the completion of the data reduction and differential photometry. This process is interactive, making it intuitive to choosing reference stars.

\subsection{Planet Transit Model fitting} \label{sec:fitting}
We performed data fitting of transit data using the \texttt{EXOFASTv2} online service, supported by NASA Exoplanet Science Institute \citep{2013PASP..125...83E,2017ascl.soft10003E}. In order to efficiently identify potential planets from a vast amount of observed TOIs transit data, we initially visually detect distinguishable transits after generating light curves and prioritize multiband light curve fitting for these targets. However, TOIs with transit depths below the photometry accuracy threshold, their transits are necessitating phase folding in an attempt to uncover the transit signal. Following data reduction and flux normalization, we input each band transit photometric data and prior distribution of parameters(see Table \ref{tab:transit_prior}) into \texttt{EXOFASTv2}, which employs MCMC analysis to perform model fitting and parameter analysis of both the host star and planet. The independent draws of MCMC and convergence criteria are referred to \citep{2013PASP..125...83E}.

\begin{table}
  \centering
  \caption{The prior distribution of transit-only fit}\label{tab:transit_prior}
\begin{tabular}{c c c c}
  \hline
  Parameter& Description(Unit) & TOI-1194 & TOI-1251 \\
  \hline
  [Fe/H]& Metallicity(dex)$^1$ & $\mathcal{N}$(0.28,0.05) & $\mathcal{N}$(-0.24,0.05) \\
  $T_{eff}$ & Effective temperature(K)$^1$ & $\mathcal{N}$(5553.93,140) & $\mathcal{N}$(6038.91,140.00) \\
  log($g$)& Surface gravity(cgs)$^1$ & $\mathcal{N}$(4.5,0.1) & $\mathcal{N}$(4.2,0.1) \\
  $T_c$ & Time of conjunction({\rm $BJD_{TDB}$})$^2$ & $\mathcal{N}$(2459632.287997,0.0001554) & $\mathcal{N}$(2459759.178675,0.0001791) \\
  $R_p/R_*$ & Ratio of Planet to Stellar Radius & $\mathcal{N}$(0.1,0.05) & $\mathcal{N}$(0.006,0.003)\\
  $i$ & Inclination(Degree) & $\mathcal{N}$(90,20) & $\mathcal{N}$(90,20) \\
  P & Planetary orbital period(days)$^2$ & $\mathcal{N}$(2.3106444,0.0000007) & $\mathcal{N}$(5.9630544,0.0000015)\\
  \hline
\end{tabular}
\begin{flushleft}
$^1$\ Starhorse \citep{2019A&A...628A..94A,2022A&A...658A..91A}. \\[0.25ex]
$^2$\ ExoFOP-TESS, \url{https://exofop.ipac.caltech.edu/tess/}
\end{flushleft}
\end{table}

\section{Verification} \label{sec:Verify}

\subsection{Multiband Light Curves Fitting} \label{sec:re-follow}
\begin{figure}[htb]
\centering
\includegraphics[width=16cm, height=22cm]{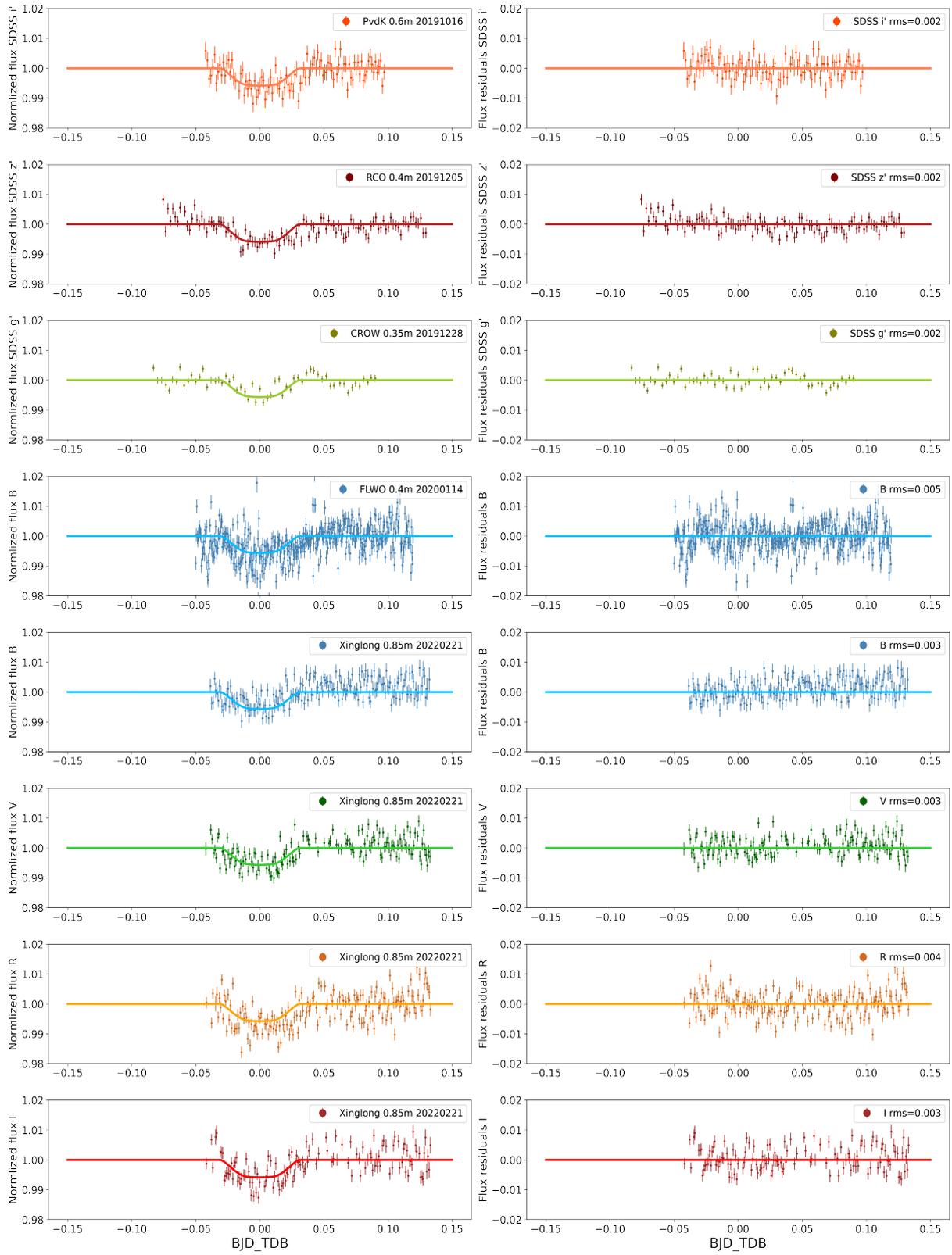}
\caption{TOI-1194 b multiband light curves with fitting models.}
 \label{1194BVRI}
\end{figure}

\begin{figure}[htb]
\centering
\includegraphics[width=16cm, height=20cm]{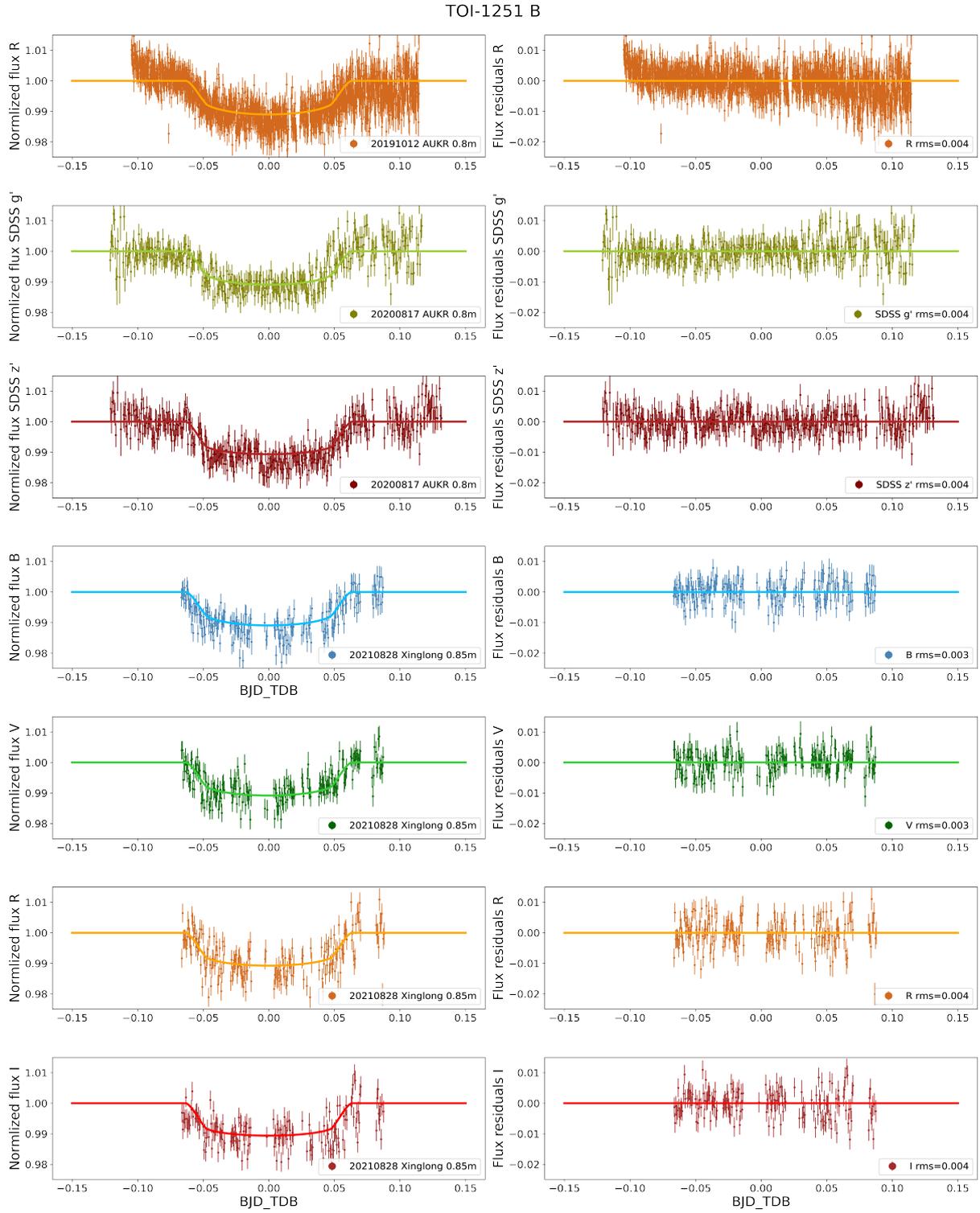}
\caption{TOI-1251 B multiband light curves with fitting models.}
 \label{1251BVRI}
\end{figure}

Comparative analysis of multiband light curves is the core of planet identification in this work. Prior to this work, detailed stellar parameters were lacking for both the host stars of TOI-1194 b and TOI-1251 B, and were therefore supplemented based on the analysis of transit photometric data and high-resolution spectra. The best-fit models for the two targets were obtained by reducing and analyzing TOI-1194 observation data obtained of Feb. $21^{\rm{st}}$, 2022 and TOI-1251 observation data obtained on Aug. $28^{\rm{th}}$, 2021. Remaining observation data were fitted using these best-fit models(see Figure \ref{1194BVRI} and \ref{1251BVRI}). The transit depth ($\delta$) and limb-darkening coefficients($u_1$, $u_2$) obtained by fitting the transit light curve of each band are shown in Table \ref{tab:band_fit}. Both candidates are achromatic in the measurement accuracy of better than 1\%.

\begin{table}
  \centering
  \renewcommand\arraystretch{1.0}
  \caption{Fitted transit depth and limb-darkening coefficient of each band}\label{tab:band_fit}
\begin{tabular}{c c c c}
  \hline
  Parameter& Description(Unit) & TOI-1194 & TOI-1251 \\
  \hline
  \textbf{Transit Depth}&  &  &  \\[0.25ex]
  ${\delta}_B$ &Transit depth of Johnson-Cousins B band (ppm) & $6684^{+80}_{-79}$ & $9975^{+358}_{-356}$ \\
  ${\delta}_V$ &Transit depth of Johnson-Cousins V band (ppm) & $6694^{+82}_{-81}$ & $9759^{+349}_{-349}$\\
  ${\delta}_R$ &Transit depth of Johnson-Cousins R band (ppm) & $6700^{+81}_{-80}$ & $10370^{+291}_{-293}$\\
  ${\delta}_I$ &Transit depth of Johnson-Cousins I band (ppm) & $6701^{+78}_{-79}$ & $9620^{+354}_{-351}$\\
  ${\delta}_{g^\prime}$ &Transit depth of SDSS $g^\prime$ band (ppm)& $6691^{+80}_{-79}$ & $10178^{+321}_{-328}$ \\
  ${\delta}_{r^\prime}$ &Transit depth of SDSS $r^\prime$ band (ppm)& ... & $10017^{+370}_{-356}$ \\
  ${\delta}_{i^\prime}$ &Transit depth of SDSS $i^\prime$ band (ppm)& $6693^{+82}_{-79}$ & $9804^{+395}_{-388}$ \\
  ${\delta}_{z^\prime}$ &Transit depth of SDSS $z^\prime$ band (ppm)& $6690\pm80$ & $10748^{+306}_{-304}$ \\
  \textbf{Limb-darkening Coefficient}&  &  & \\[0.25ex]
  $u_1$(B) & Linear limb-darkening coeff of Johnson-Cousins B band & $0.781\pm0.051$ & $0.667_{-0.051}^{+0.050}$ \\
  $u_2$(B) & Quadratic limb-darkening coeff of Johnson-Cousins B band& $0.061\pm0.050$ & $0.158\pm0.050$ \\
  $u_1$(V) & Linear limb-darkening coeff of Johnson-Cousins V band& $0.563_{-0.052}^{+0.051}$ & $0.459\pm0.051$ \\
  $u_2$(V) & Quadratic limb-darkening coeff of Johnson-Cousins V band& $0.192\pm0.050$ & $0.256_{-0.049}^{+0.050}$ \\
  $u_1$(R) & Linear limb-darkening coeff of Johnson-Cousins R band& $0.450_{-0.052}^{+0.050}$ & $0.370_{-0.049}^{+0.051}$ \\
  $u_2$(R) & Quadratic limb-darkening coeff of Johnson-Cousins R band& $0.231_{-0.049}^{+0.051}$ & $0.282_{-0.050}^{+0.048}$ \\
  $u_1$(I) & Linear limb-darkening coeff of Johnson-Cousins I band& $0.349_{-0.050}^{+0.051}$ & $0.278_{-0.050}^{+0.049}$ \\
  $u_2$(I) & Quadratic limb-darkening coeff of Johnson-Cousins I band& $0.247_{-0.049}^{+0.050}$ & $0.268_{-0.049}^{+0.051}$ \\
  $u_1$($g^\prime$) &Linear limb-darkening coeff of SDSS $g^\prime$ band& $0.696\pm0.051$ & $0.585\pm0.051$ \\
  $u_2$($g^\prime$) &Quadratic limb-darkening coeff of SDSS $g^\prime$ band& $0.119\pm0.051$ & $0.221\pm0.050$ \\
  $u_1$($r^\prime$) &Linear limb-darkening coeff of SDSS $r^\prime$ band& $ ... $ & $0.386^{+0.049}_{-0.051}$ \\
  $u_2$($r^\prime$) &Quadratic limb-darkening coeff of SDSS $r^\prime$ band& ... & $0.277^{+0.051}_{-0.050}$ \\
  $u_1$($i^\prime$) &Linear limb-darkening coeff of SDSS $i^\prime$ band& $0.377\pm0.050$ & $0.316\pm0.050$ \\
  $u_2$($i^\prime$) &Quadratic limb-darkening coeff of SDSS $i^\prime$ band& $0.246\pm0.050$ & $0.282^{+0.050}_{-0.051}$ \\
  $u_1$($z^\prime$) &Linear limb-darkening coeff of SDSS $z^\prime$ band& $0.304\pm0.050$ & $0.273^{+0.050}_{-0.051}$ \\
  $u_2$($z^\prime$) &Quadratic limb-darkening coeff of SDSS $z^\prime$ band & $0.254\pm0.050$ & $0.291^{+0.049}_{-0.050}$ \\
  \hline
\end{tabular}
\end{table}

\subsection{Radial Velocity Fitting}\label{sec:RV_fit}
Determining the mass of the companions enables direct confirmation of their planetary status. To accomplish this, we reduced the high-resolution spectra of TOI-1194 and TOI-1251 captured by TRES, FIES, and HRS instrument, calculated their radial velocities and derive their phase-RV variation curves, before analyzing their orbital solutions. Reduction and multi-order spectrum analysis from TRES and FIES were conducted using the method described in \citep{2010ApJ...720.1118B}. We utilized Stellar Parameters Classification tool(SPC) to estimate the stellar parameters. The SPC utilizes synthetic grid library spectra and simultaneously derive the effective temperature, surface gravity, metallicity, and rotational velocity by matching the models with the observed spectra originating from different instruments \citep{2012Natur.486..375B,2014Natur.509..593B}(The reference stellar atmospheric models from \citep{1992IAUS..149..225K}). High-resolution RV data obtained from follow-up observations by HRS was reduced using \texttt{GAMSE}\footnote{\url{https://github.com/wangleon/gamse}} for 1-d spectra extraction from double-fiber high-resolution spectra. We employed \texttt{IRAF} to fit nine lines in Order93 and Order98 for calculating RV in the 600-700 nm \citep{1986SPIE..627..733T}. We utilized \texttt{barycorrpy} \citep{2018ascl.soft08001K} to calculated the barycentric RV correction. Three points of HRS RV data obtained on Nov. $22^{\rm{nd}}$, 2021 were disregarded, due to poor SNRs and insufficient identifiable lines. To enable fitting of all RV points collected from the three instruments, we shifted the systemic velocity($\gamma$) of each instrument to 0 m/s. Table \ref{tab:TOI1194_rv} shows the RV data of TOI-1194 b, while Table \ref{tab:TOI1251_rv} shows RV data of TOI-1251 B, within best models obtained when the$\chi^{2}/dof$ value is lowest.

\begin{figure}[htb]
\centering
\includegraphics[width=7.5cm, height=6.5cm]{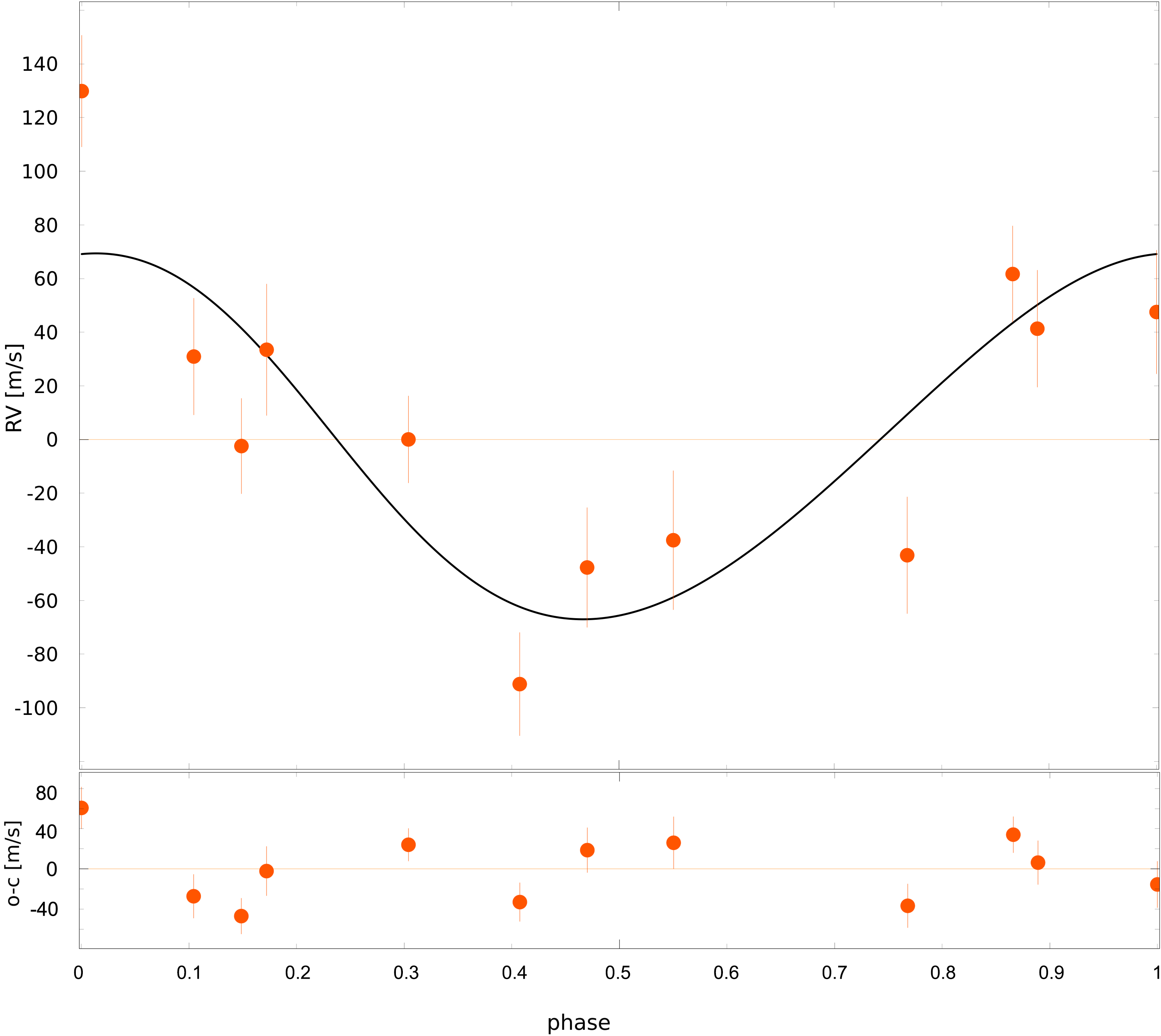}
\caption{The fitted RV curve of TOI-1194 b by follow-up high resolution spectra.}
 \label{1194rvcurve}
\end{figure}

\begin{figure}[htb]
\centering
\includegraphics[width=8cm, height=7cm]{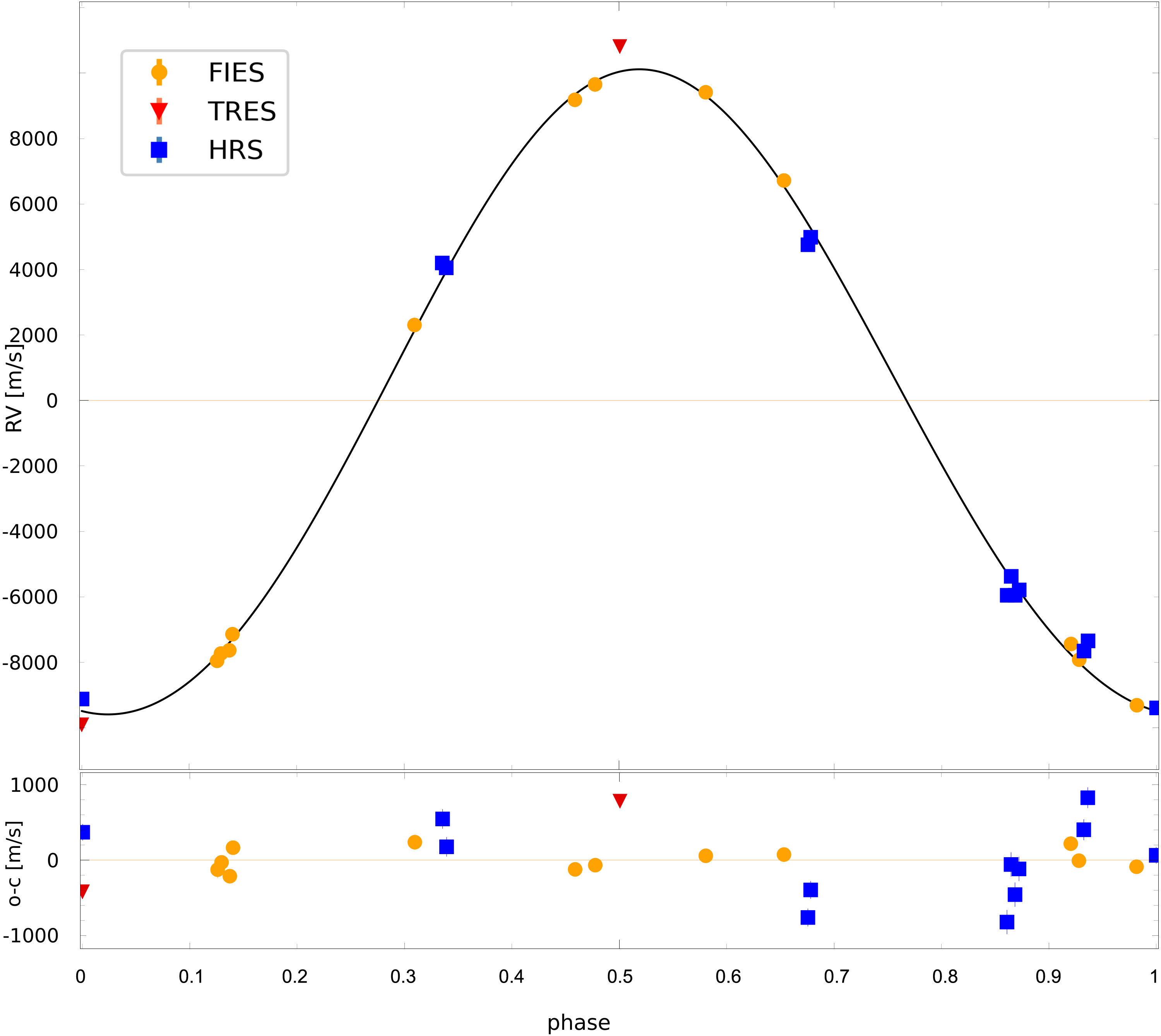}
\caption{The fitted RV curve of TOI-1251 B by follow-up high resolution spectra.}
 \label{1251rvcurve}
\end{figure}

We utilized both \texttt{EXOFASTv2} and \texttt{Exo-Striker}\citep{2019ascl.soft06004T} to fit RV points for two TOIs, with best-fitted models visualized in Figure \ref{1194rvcurve} and \ref{1251rvcurve}. The calculated RV amplitude caused by TOI-1194 b to be K = $69.4_{-7.3}^{+7.9}$ m/s, derived a minimum mass $M_{p}$sin$i$=$0.453_{-0.051}^{+0.055} M_J$, closely related to Saturn in terms of mass and orbiting its host star on a near-circular orbit.

The RV amplitude of TOI-1251 is notably distinct different from that of TOI-1194, duo to the former is a very low-mass star. We calculate the RV amplitude caused by TOI-1251 B to be K = $9849_{-40}^{+42}$ m/s, with a minimum mass of $M_2$sin$i$ = $96.3\pm1.5$ $M_J$. The fitted RV curve suggests that TOI-1251 B orbits the host star on a near-circular orbit, consistent with parameters obtained in fitted light curves.

\begin{table}[htb]
\centering
\caption{Ground-Based Follow-Up Radial Velocities for TOI-1194 b \label{tab:TOI1194_rv}}
\begin{tabular}{c c c c}
  \hline
  \bjdtdb & RV & \textup{$e_{RV}$} & Instrument\\
   & (m/s) & (m/s) & \\
  \hline
  2458829.952974 &  129.80 & 20.77 & FLWO 1.5m TRES\\
  2458830.979613 & -47.76  & 22.29 & FLWO 1.5m TRES\\
  2459980.881301 &  30.22  & 20.77 & FLWO 1.5m TRES\\
  2459981.855344 & -38.24  & 25.88 & FLWO 1.5m TRES\\
  2459982.837025 &  46.81  & 23.01 & FLWO 1.5m TRES\\
  2459983.853953 & -91.88  & 19.20 & FLWO 1.5m TRES\\
  2459984.855371 &  60.96  & 17.91 & FLWO 1.5m TRES\\
  2459985.938541 &  -0.69  & 16.18 & FLWO 1.5m TRES\\
  2459987.909966 &  -3.17  & 17.74 & FLWO 1.5m TRES\\
  2459991.837408 &  40.57  & 21.77 & FLWO 1.5m TRES\\
  2460000.815599 & -43.86  & 21.72 & FLWO 1.5m TRES\\
  2460001.824999 &  32.74  & 24.48 & FLWO 1.5m TRES\\
  \hline
\end{tabular}
\end{table}

\begin{table}[htb]
\centering
\caption{Ground-Based Follow-Up Radial Velocities for TOI-1251 B \label{tab:TOI1251_rv}}
\begin{tabular}{c c c c}
  \hline
  \bjdtdb & RV & \textup{$e_{RV}$} & Instrument\\
   & (m/s) & (m/s)& \\
  \hline
  2458776.68854 & -10363.24 &  67.18 & FLWO 1.5m TRES\\
  2458779.66325 &  10356.45 &  67.18 & FLWO 1.5m TRES\\
  2459026.64331 &  -7527.15 &  67.48 & NOT 2.56m FIES\\
  2459036.54954 &   9326.95 &  91.73 & NOT 2.56m FIES\\
  2459038.61436 &  -8007.87 &  53.32 & NOT 2.56m FIES\\
  2459093.47847 &  -8041.56 & 108.67 & NOT 2.56m FIES\\
  2459093.50021 &  -7818.58 &  79.84 & NOT 2.56m FIES\\
  2459095.45703 &   9096.31 &  56.86 & NOT 2.56m FIES\\
  2459105.47225 &  -7720.50 &  72.90 & NOT 2.56m FIES\\
  2459119.42068 &   9566.68 &  62.45 & NOT 2.56m FIES\\
  2459123.37912 &  -7229.17 &  83.26 & NOT 2.56m FIES\\
  2459124.38530 &   2216.98 &  98.45 & NOT 2.56m FIES\\
  2459132.39111 &   6631.18 &  52.06 & NOT 2.56m FIES\\
  2459134.34215 &  -9398.61 &  50.26 & NOT 2.56m FIES\\
  2459539.93873  &  -9629.71 &  107.67 & Xinglong 2.16m HRS \\
  2459539.96100  &  -9357.86 &  107.67 & Xinglong 2.16m HRS \\
  2459631.39846  &  3962.04 &  126.91 & Xinglong 2.16m HRS \\
  2459631.42052  &  3814.53 &  126.91 & Xinglong 2.16m HRS \\
  2459639.38316  &  4519.53 &  116.19 & Xinglong 2.16m HRS \\
  2459639.39846  &  4748.09 &  116.19 & Xinglong 2.16m HRS \\
  2459754.20779  & -7892.08 &  135.92 & Xinglong 2.16m HRS \\
  2459754.22996  & -7582.42 &  135.92 & Xinglong 2.16m HRS \\
  2459890.93302  & -6189.59 &  158.63 & Xinglong 2.16m HRS \\
  2459890.95543  & -5610.81 &  158.63 & Xinglong 2.16m HRS \\
  2459890.97752  & -6190.05 &  158.63 & Xinglong 2.16m HRS \\
  2459890.99957  & -6025.26 &  158.63 & Xinglong 2.16m HRS \\
  \hline
\end{tabular}
\end{table}

\subsection{Joint Analysis}
The parameters of the TOI-1194 b and TOI-1251 B were derived through a joint analysis of all available ground-based photometric and spectroscopic data. We employed Exo-Striker to conduct the joint MCMC analysis\citep{2013PASP..125..306F}, initializing 20 walkers and generating 2000 steps for the burn-in phase, further 10000 steps to sample the posterior parameters distribution of stellar and planetary parameters from the MCMC process.

In addition to the transit epoch ($T_c$) and period (P) in Table \ref{tab:transit_prior}, the prior parameters used also include metallicity ([Fe/H]), surface gravity (log$g$), and effective temperature ($T_{eff}$), which were obtained by fitting stellar spectra (Section \ref{sec:RV_fit}). Both planets appear to be in near-circular orbits based on the RV curve fitting, we assigned a Gaussian distribution $\mathcal{N}$(0,0.1) as a prior.

The process ultimately converged to the maximum likelihood estimate. The joint analysis results, presented at 1$\sigma$ confidence interval in Table \ref{tab:1194par} and \ref{tab:1251par}, indicate that the mass of TOI-1194 b is $0.456_{-0.051}^{+0.055} M_J$, while the mass of TOI-1251 B is $96.5\pm1.5 M_J$.

\begin{table*}[htb]
\scriptsize
\setlength{\tabcolsep}{2pt}
\centering
\renewcommand\arraystretch{0.8}
\caption{TOI-1194 host stellar and planetary parameters at 68\% confidence intervals}
\label{tab:1194par}
\begin{tabular*}{\textwidth}{l @{\extracolsep{\fill}} lcccc}
  \hline
  \hline
Parameter & Description (Units) & \multicolumn{4}{c}{Values} \\
\hline
\multicolumn{2}{l}{\textbf{Host Stellar Parameters}}\\
~~~~$RA$\dotfill &RA coordinate (J2000.0)$^{1}$ \dotfill & $11h11m16.91s$\\
~~~~$Dec$\dotfill &Dec coordinate (J2000.0)$^{1}$ \dotfill & $+69d57m52.94s$\\
~~~~${\rm B [mag]}$\dotfill &Johnson-Cousins B band magnitude$^{2}$ \dotfill & $11.833\pm0.081$\\
~~~~${\rm V [mag]}$\dotfill &Johnson-Cousins V band magnitude$^{3}$ \dotfill & $11.185\pm0.042$\\
~~~~${\rm G [mag]}$\dotfill &Gaia G band magnitude$^{4}$ \dotfill & $10.9887\pm0.0004$\\
~~~~${\rm I [mag]}$\dotfill &Johnson-Cousins I band magnitude$^{3}$ \dotfill & $10.289\pm0.020$\\
~~~~$M_*$\dotfill &Mass (\msun)\dotfill & $1.007_{-0.048}^{+0.050}$\\
~~~~$R_*$\dotfill &Radius (\rsun)\dotfill & $0.963_{-0.051}^{+0.056}$\\
~~~~$L_*$\dotfill &Luminosity (\lsun)\dotfill & $0.732_{-0.081}^{+0.095}$\\
~~~~$\rho_*$\dotfill &Density (cgs)\dotfill & $1.59_{-0.22}^{+0.26}$\\
~~~~$\log{g}$\dotfill &Surface gravity (cgs)\dotfill & $4.473_{-0.042}^{+0.043}$\\
~~~~$T_{\rm TIC}$\dotfill &Effective Temperature from TIC(K)$^{5}$\dotfill &$5323.0^{+92.9}_{-156.1}$\\
~~~~$T_{\rm Gaia}$\dotfill &Effective Temperature from Gaia (K)$^{4}$\dotfill &$5339.90^{+207.10}_{-116.23}$\\
~~~~$T_{\rm eff}$\dotfill &Best fitting Effective Temperature (K)\dotfill & $5446^{+51}_{-48}$\\
~~~~$[{\rm Fe/H}]$\dotfill &Metallicity (dex)\dotfill & $0.290\pm0.078$\\
~~~~$v${\rm sin}$i$\dotfill &Rotational velocity (km/s)\dotfill & $2.4\pm0.5$\\
~~~~$\varpi$\dotfill &Parallax (mas)$^{1}$\dotfill &$6.6527\pm0.0126$\\
~~~~$d$\dotfill &Distance (kpc)$^{1}$\dotfill &$0.148508^{+0.003023}_{-0.003278}$\\
[1ex]
\hline
\multicolumn{2}{l}{\textbf{Planetary Companion Parameters}} & TOI-1194 b\\
~~~~$P$\dotfill &Period (days)\dotfill & $2.310644\pm0.000001$\\
~~~~$R_P$\dotfill &Radius (\rj)\dotfill & $0.767_{-0.041}^{+0.045}$\\
~~~~$M_{P}$\dotfill &Mass (\mj)\dotfill & $0.456_{-0.051}^{+0.055}$\\
~~~~$\rho_{P}$\dotfill &Density (cgs)\dotfill & $1.25_{-0.21}^{+0.26}$\\
~~~$\log(g_{_P})$\dotfill &Surface gravity\dotfill & $3.281\pm0.064$\\
~~~~$M_{P}/M_*$\dotfill &Mass ratio \dotfill & $0.000432_{-0.000046}^{+0.000050}$\\
~~~~$R_P/R_*$\dotfill &Radius of planet in stellar radii \dotfill & $0.08196\pm0.00047$\\
~~~~$a/R_*$\dotfill &Semi-Major Axis in Stellar Radii \dotfill & $7.65_{-0.37}^{+0.39}$\\
~~~~$\delta$\dotfill &Transit depth (fraction) \dotfill & $0.006717\pm0.000078$\\
~~~~$K$\dotfill &RV Semi-Amplitude (m/s) \dotfill & $69.4_{-7.3}^{+7.9}$\\
~~~~$T_0$\dotfill &Time of conjunction (\bjdtdb)\dotfill & $2459632.28801\pm0.00015$\\
~~~~$a$\dotfill &Semi-major axis (AU)\dotfill & $0.03428\pm0.00055$\\
~~~~$i$\dotfill &Inclination (Degrees)\dotfill & $83.98\pm0.50$\\
~~~~$b$\dotfill &Transit Impact parameter \dotfill & $0.802_{-0.027}^{+0.025}$\\
~~~~$e$\dotfill &Eccentricity\dotfill &$0.076_{-0.043}^{+0.032}$\\
~~~~$\omega$\dotfill &Argument of Periastron (Degree) \dotfill & $76.7_{-39.5}^{+49.7}$\\
~~~~$T_{14}$\dotfill &Total transit duration (hours)\dotfill &$1.684^{+0.063}_{-0.058}$\\
~~~~$T_{eq}$\dotfill &Equilibrium temperature (K)\dotfill & $1391_{-37}^{+38}$\\
~~~~\textup{$\chi_{RV}^{2}$}\dotfill &RV fitting $\chi^{2}$ value \dotfill & 3.89\\
~~~~\textup{$rms_{RV}$}\dotfill &RV fitting rms value (m/s)\dotfill & 31.8\\
[1ex]
\hline \\[-6ex]
\end{tabular*}
\begin{flushleft}
\footnotesize{\vspace{6pt}{\bf Note.}\\[0.25ex]
$^1$\ Gaia DR3 \citep{2023A&A...674A...1G}.\\[0.25ex]
$^2$\ All-sky compiled catalogue of 2.5 million stars (ASCC-2.5 V3) \citep{2001KFNT...17..409K}.\\[0.25ex]
$^3$\ TASS Mark IV Photometric Survey of the Northern Sky \citep{2006PASP..118.1666D} \\[0.25ex]
$^4$\ Gaia DR2 \citep{2018yCat.1345....0G}. \\[0.25ex]
$^5$\ TIC v8.2 \citep{2021arXiv210804778P}. \\[0.25ex]
}
\end{flushleft}
\end{table*}

\begin{table*}[htb]
\scriptsize
\setlength{\tabcolsep}{2pt}
\centering
\renewcommand\arraystretch{0.8}
\caption{TOI-1251 Host and Companion Stellar Parameters at 68\% Confidence Intervals}
\label{tab:1251par}
\begin{tabular*}{\textwidth}{l @{\extracolsep{\fill}} lcccc}
  \hline
  \hline
Parameter & Description (Units) & \multicolumn{4}{c}{Values} \\
\hline
\multicolumn{2}{l}{\textbf{Host Stellar Parameters}}\\
~~~~$RA$\dotfill &RA coordinate (J2000.0)$^{1}$ \dotfill & $18h14m07.07s$\\
~~~~$Dec$\dotfill &Dec coordinate (J2000.0)$^{1}$ \dotfill & $+62d51m29.71s$\\
~~~~${\rm B [mag]}$\dotfill &Johnson-Cousins B band magnitude$^{2}$ \dotfill & $12.020\pm0.127$\\
~~~~${\rm V [mag]}$\dotfill &Johnson-Cousins V band magnitude$^{3}$ \dotfill & $11.318\pm0.037$\\
~~~~${\rm G [mag]}$\dotfill &Gaia G band magnitude$^{4}$ \dotfill & $11.0942\pm0.0010$\\
~~~~${\rm I [mag]}$\dotfill &Johnson-Cousins I band magnitude$^{3}$ \dotfill & $10.505\pm0.021$\\
~~~~$M_*$\dotfill &Mass (\msun)\dotfill & $1.057_{-0.049}^{+0.050}$\\
~~~~$R_*$\dotfill &Radius (\rsun)\dotfill & $0.986_{-0.024}^{+0.025}$\\
~~~~$L_*$\dotfill &Luminosity (\lsun)\dotfill & $0.968_{-0.062}^{+0.066}$\\
~~~~$\rho_*$\dotfill &Density (cgs)\dotfill & $1.553_{-0.059}^{+0.061}$\\
~~~~$\log{g}$\dotfill &Surface gravity (cgs)\dotfill & $4.4737_{-0.0093}^{+0.0096}$\\
~~~~$T_{\rm Gaia}$\dotfill &Effective Temperature from Gaia (K)$^{4}$\dotfill &$5759.71^{+163.69}_{-209.53}$\\
~~~~$T_{\rm eff}$\dotfill &Best fitting Effective Temperature (K)\dotfill & $5769_{-50}^{+49}$\\
~~~~$[{\rm Fe/H}]$\dotfill &Metallicity (dex) \dotfill & $0.019\pm0.080$\\
~~~~$v${\rm sin}$i$\dotfill &Rotational velocity (km/s)\dotfill & $9.2\pm0.5$\\
~~~~$\varpi$\dotfill &Parallax (mas)$^{1}$\dotfill & $3.6550\pm0.3382$\\
~~~~$d$\dotfill &Distance (kpc)$^{1}$\dotfill & $0.246147_{-0.011126}^{+0.010750}$\\
[1ex]
\hline
\multicolumn{2}{l}{\textbf{Companion Stellar Parameters}} & TOI-1251 B\\
~~~~$P$\dotfill &Period (days)\dotfill & $5.963054\pm0.000001$\\
~~~~$R_2$\dotfill &Radius (\rj)\dotfill & $0.947_{-0.035}^{+0.037}$\\
~~~~$M_2$\dotfill &Mass (\mj)\dotfill & $96.47_{-1.47}^{+1.50}$\\
~~~$\rho_{2}$\dotfill &Density (cgs)\dotfill & $141_{-12}^{+13}$\\
~~~$\log(g_{2})$\dotfill &Surface gravity\dotfill & $5.425_{-0.025}^{+0.024}$\\
~~~~$M_2/M_*$\dotfill &Mass ratio \dotfill & $0.0876_{-0.0008}^{+0.0007}$\\
~~~~$R_2/R_*$\dotfill &Radius of planet in stellar radii \dotfill & $0.0987\pm0.0022$\\
~~~~$a/R_*$\dotfill &Semi-Major Axis in Stellar Radii \dotfill & $14.69\pm0.19$\\
~~~~$\delta$\dotfill &Transit depth (fraction) \dotfill & $0.00975_{-0.00042}^{+0.00043}$\\
~~~~$K$\dotfill &RV Semi-Amplitude (m/s) \dotfill & $9849_{-40}^{+42}$\\
~~~~$T_0$\dotfill &Time of conjunction (\bjdtdb)\dotfill & $2459759.17874_{-0.00018}^{+0.00017}$\\
~~~~$a$\dotfill &Semi-major axis (AU)\dotfill & $0.0674\pm0.0010$\\
~~~~$i$\dotfill &Inclination (Degrees)\dotfill & $92.3_{-0.11}^{+0.08}$\\
~~~~$b$\dotfill &Transit Impact parameter \dotfill & $0.469_{-0.030}^{+0.027}$\\
~~~~$e$\dotfill &Eccentricity \dotfill & $0.028_{-0.001}^{+0.002}$\\
~~~~$\omega$\dotfill &Argument of Periastron (Degree) \dotfill & $341_{-5}^{+6}$\\
~~~~$T_{14}$\dotfill &Total transit duration (hours)\dotfill & $3.086_{-0.040}^{+0.041}$\\
~~~~\textup{$\chi_{RV}^{2}$}\dotfill &RV fitting $\chi^{2}$ value \dotfill & 3.37\\
~~~~\textup{$rms_{RV}$}\dotfill &RV fitting rms value (m/s)\dotfill & 390.7\\
[1ex]
\hline \\[-6ex]
\end{tabular*}
\begin{flushleft}
\footnotesize{\vspace{6pt}{\bf Note.}\\[0.25ex]
$^1$\ Gaia DR3 \citep{2023A&A...674A...1G}.\\[0.25ex]
$^2$\ All-sky compiled catalogue of 2.5 million stars (ASCC-2.5 V3) \citep{2001KFNT...17..409K}.\\[0.25ex]
$^3$\ TASS Mark IV Photometric Survey of the Northern Sky \citep{2006PASP..118.1666D} \\[0.25ex]
$^4$\ Gaia DR2 \citep{2018yCat.1345....0G}. \\[0.25ex]
}
\end{flushleft}
\end{table*}

\subsection{Potential Contamination} \label{sec:contamination}
\begin{figure}[htb]
\centering
\includegraphics[width=8cm, height=6cm]{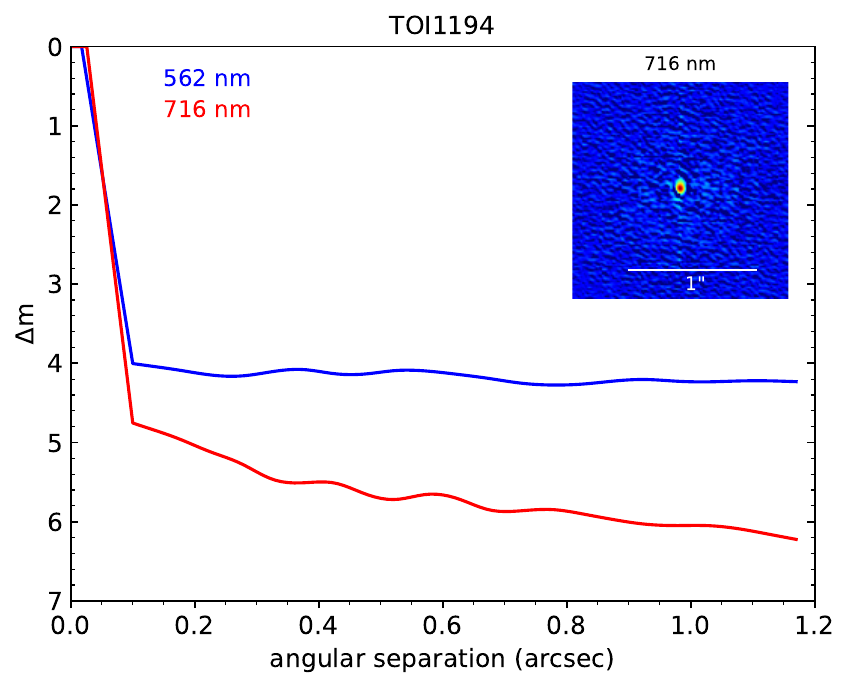}
\caption{The blue and red curves show the 5$\sigma$ contrast sensitivity limits in 562 and 716 nm respectively of TOI-1194. The observations were obtained using the 'Alopeke instrument on the Gemini North telescope. The upper right inset shows the reconstructed 716 nm speckle image.}
 \label{1194speckle}
\end{figure}

Apart from the EBs mentioned in Section\textcolor{red}{.} \ref{sec:intro}, the eclipses of nearby EBs are usually responsible for the high FP possibility in exoplanet surveys as well. Approximately half of all discovered planets orbit stars with stellar companions \citep{2021AAS...23734401L,2021AJ....161..164H}. However, follow-up photometric data and Gaia DR3 catalogs alone are not sufficient to distinguish such companions in regions between 10-100 AU around the host stars. High-resolution imaging data have the potential to exclude some companions around the host stars \citep{2021AJ....162...75L}. The resolution of ground-based optical telescopes without adaptive optics depends on the atmosphere seeing conditions. At the Xinglong Observatory, the median seeing value is around 1.7\arcsec \citep{2015PASP..127.1292Z}, incapable to rule out stellar contaminations from stars present in this sky region via aperture photometry alone. We used the 'Alopeke instrument on Gemini North 8-m telescope to observe TOI-1194 on Feb. 17$^{th}$, 2020 and obtained high-resolution speckle images in the 562nm and 716nm bands to search for nearby stellar companions (see Figure \ref{1194speckle}). The data were reduced using the standard speckle interferometry pipeline as described in \citep{2011AJ....142...19H}. From the 5$\sigma$ contrast curves shown in Figure. \ref{1194speckle}, we found that within the angular and magnitude limits achieved, there is no close companion star within 5-7 magnitudes and between 0.02\arcsec(the diffraction limit of device) to 1\arcsec. These angular limits correspond to 2.96 AU to 148 AU around TOI-1194, which is located 148 pc away.

The 562nm and 832nm speckle images of TOI-1251 from WIYN telescope revealed the presence of a star located 0.22\arcsec away from the host star (equivalent to a projected separation of about 41 AU), with $\Delta$Mag = 1.66 in 562nm band and $\Delta$Mag = 1.36 in 832nm band\citep{2021AJ....161..164H}. However, from follow-up photometric and spectroscopic radial velocity data, we discerned the existence of only one stellar companion, which orbits host star in a near-circular orbit with a semi-major axis of a = $0.0674\pm0.0010$AU. While the angular separation between the companion and the host star observed from the Earth's perspective is 0.0002\arcsec, high-resolution imaging cannot currently observe it. This in turn indicates that the star located within 41 AU of the host star is not dynamically associated with it.

\section{Summary and Discussion} \label{sec:sum}
In our efforts to follow-up on TOIs, we organized a Sino-German campaign group collected multiple transit photometric data observed in multiband to validate them as planets. The first three years of observation and current data analysis demonstrate that the one meter-class telescopes in our campaign can provide valuable transit and RV data for the identification and planetary properties analysis of TESS candidates. Through processing and fitting of these data, we confirmed the present of a planet, TOI-1194 b, and a very low-mass star, TOI-1251 B, each orbiting G-type hosts. Identifying a planet using only single-band photometric observations collected over a long time baseline is quite difficult due to differences in transit depth across multiband light curves displayed by EBs. In addition to improving observational accuracy and increasing transit data availability, multiband simultaneous photometric observation data is also helpful for the rapid elimination of FP caused by EBs. Besides, one meter-class telescopes are more accessible and offer cheaper observation resources compared to spectral observation equipment. The ability to conduct multiband photometric observation of exoplanets through multi-site campaign can also enable faster response times and feedback on follow-up observation results for candidates detected through space-based observations like TESS, enhancing the efficiency and accuracy of exoplanet confirmation.

\begin{figure}[htb]
\centering
\includegraphics[width=9cm, height=6cm]{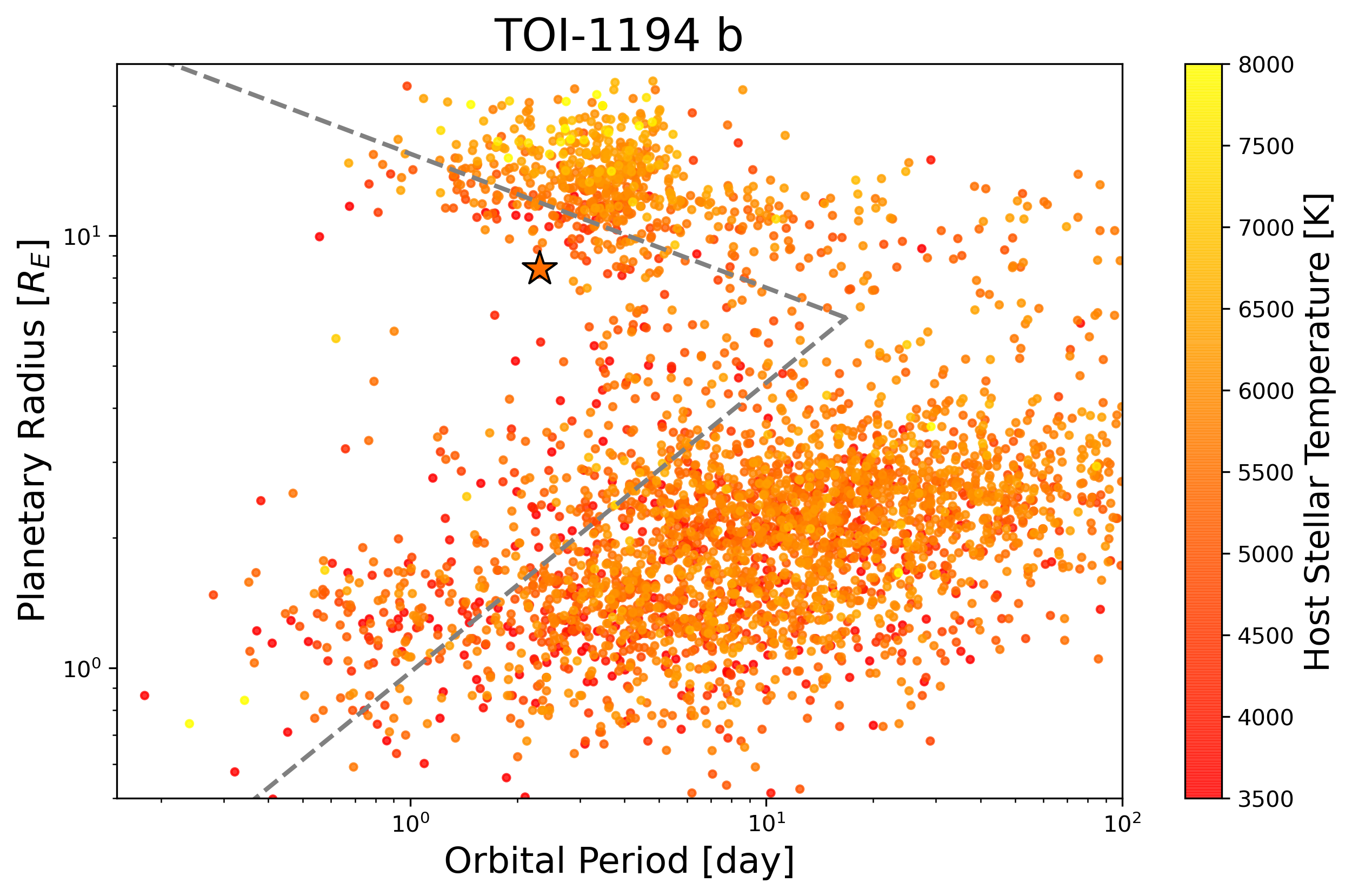}
\caption{The period-radius diagram of confirmed planets, grey dash lines are the region of Neptune desert described in \citep{2016A&A...589A..75M}}
 \label{1194PR}
\end{figure}

\begin{figure}[htb]
\centering
\includegraphics[width=8cm, height=6cm]{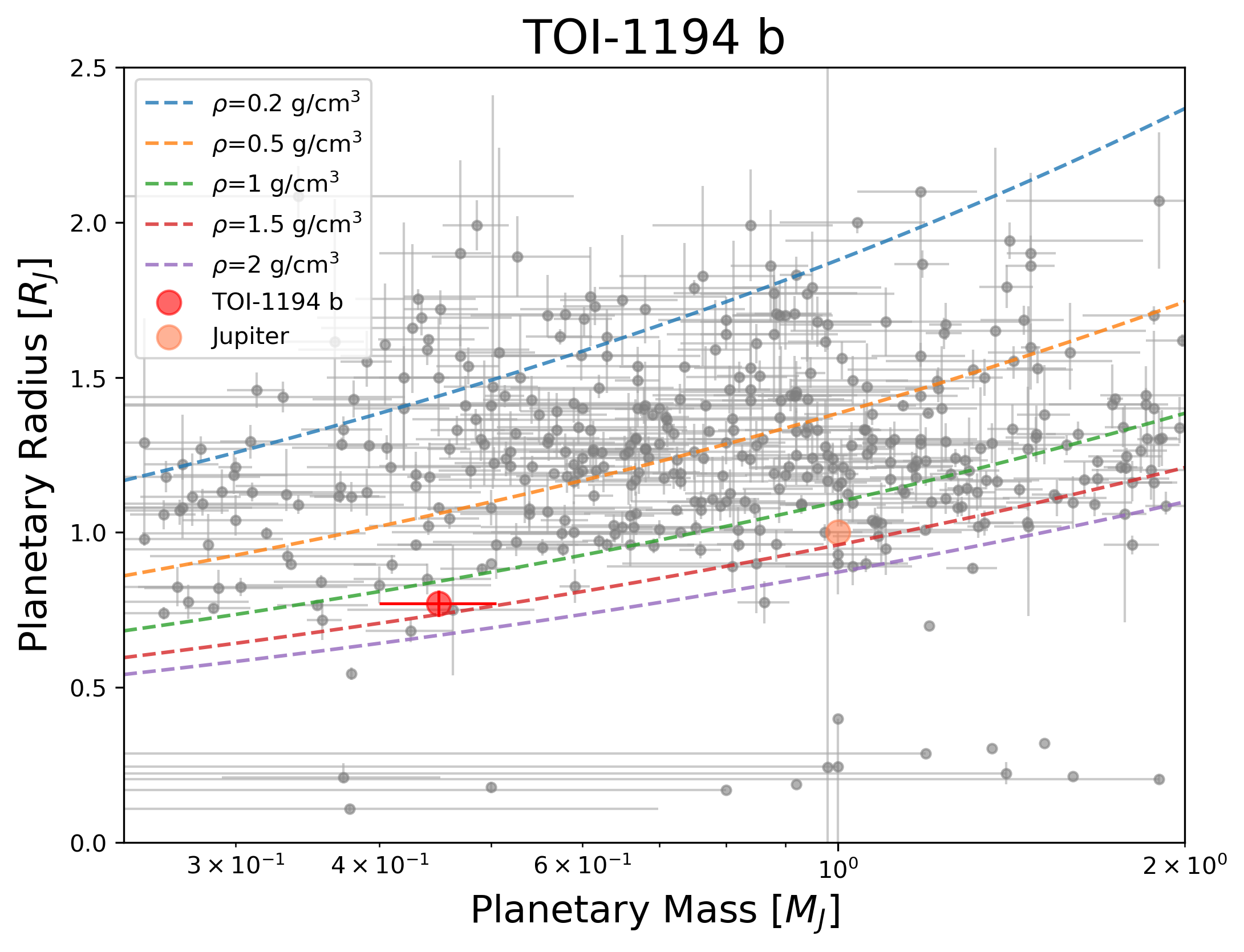}
\caption{The mass-radius diagram of confirmed hot Jupiters ($M_P$ $\geq$ 0.25 \mj, $P$ $\leq$ 10 days) with mass $\le$ 2\mj, the red point is TOI-1194 b, the yellow point is Jupiter, dash lines with different colors corresponding the density values shown in the legend. The density of TOI-1194 b is close to that of Jupiter and is relatively dense among these hot Jupiter shown here.}
 \label{1194MR}
\end{figure}

TOI-1194 b has a density of $\rho$ = $1.25_{-0.21}^{+0.26}$ $g/cm^3$, suggesting that this hot Saturn is comparable in density to Jupiter and is located in the Neptune desert in \citep{2016A&A...589A..75M}(see Figure \ref{1194PR}). Figure \ref{1194MR} illustrates the mass-radius diagram of confirmed hot Jupiters with mass under 2\mj, with TOI-1194 falling into a relative dense and sparsely-populated region\citep{2017MNRAS.471.4374E}. As statistics of ultra-short period planets (USP) samples and evolution model of hot gas giant planets, it is hypothesized that planets with this nature, exposed to strong ultraviolet and thermal radiation from their host stars, may experience expansion and gradual loss of their outer envelope over time\citep{2013ApJ...775..105O}. The radius expansion of the planet also increases with the enhanced radiation from the host star, and this radiation-induced inflation mechanism is evident at $T_{eq}$ $\ge$ 1000K\citep{2011ApJS..197...12D}. The $T_{eq}$ of TOI-1194 b is $\sim$ 1400K, but the radius does not show a significant expansion effect, so it is relatively dense. The higher density of TOI-1194 b may be attribute to its higher proportion of heavy elements\citep{2011ApJ...736L..29M}, and the relationship between planet and metal-rich host star may be reasonably explained by protoplanetary disk evolution theories.

TOI-1251 B is an M-dwarf and located remarkably close to its host star, with an extreme low mass ratio q = $M_{2}/M^{*}$=0.088, but positioned outside the Roche limit(0.00025AU). M dwarfs are widely distributed in the universe, due to the smaller radius and mass of M dwarfs, the transit depth and radial velocity variation caused by smaller planets (such as Neptune and Earth-like planets) are more pronounced than those around Sun-like stars\citep{2019arXiv190604644T}, and these small planets are more likely to be detected by one meter-class telescope, so they have attracted much attention in recent years. The statistical results demonstrate a significantly high occurrence rate of low-mass planets around M dwarfs, with a correspondingly high likelihood of super-Earths (3-10$M_{\bigoplus}$) existing within their habitable zones\citep{2014MNRAS.441.1545T,2015ApJ...807...45D}. Although most M-dwarfs identified in EB systems exhibit mass similar to that of another star in the system, there are fewer than 100 M-dwarfs orbiting F/G/K-type stars with an extremely low mass ratio (q $\le$ 0.15) \citep{2021A&A...652A.127G}. Due to the limited sample, the origin and evolution of these EBs lack a systematic explanation. TOI-1251 B expands the sample as a new low-mass eclipsing binaries (EBLM), with the period obtained from this work being consistent with that provided within the Gaia DR3 Non-single stars catalog. This consistency indicates that it is a single-lined spectroscopic binary(SB) \citep{2022yCat.1357....0G}. Direct observation
of very low mass stars are challenging due to their low surface luminosity compared to massive stars. Nonetheless, the methods employed in exoplanet follow-up observation are well-suited to the observation of these types of stars. Therefore, identifying new EBLMs in TOI candidates using follow-up photometric and spectroscopic observations may become easier.

\begin{acknowledgements}
We thank Dr. David W. Latham and the TRES team for providing the TRES results and helpful suggestions in improving the manuscript. This work is supported by National Natural Science Foundation of China U1831209, U2031144 and the research fund of Ankara University (BAP) through the project 18A0759001. This paper refer to some data which are publicly available from the Mikulski Archive for Space Telescopes (MAST), collected by the TESS mission. The TESS mission is funded by NASA's Science Mission directorate. We acknowledge the TESS Follow-up Observing Program (TFOP) SG1, SG2, SG3 and SG4 for the disposition of candidates, and the use of public TOI Release data from pipelines at the TESS Science Office(TSO) and at the TESS Science Processing Operations Center. Resources supporting this work were provided by the NASA High-End Computing (HEC) Program through the NASA Advanced Supercomputing (NAS) Division at Ames Research Center for the production of the SPOC data products. This research has made use of data obtained from the portal exoplanet.eu of The Extrasolar Planets Encyclopaedia. This research has made use of the NASA Exoplanet Archive, which is operated by the California Institute of Technology, under contract with the National Aeronautics and Space Administration under the Exoplanet Exploration Program.

This article is partly based on observations made with the FLWO/TRES, NOT/FIES and MuSCAT2. We acknowledge the support of the staff of the Xinglong 60cm, 80cm, 85cm and 2.16m telescopes. This work was partially supported by the Open Project Program of the CAS Key Laboratory of Optical Astronomy, National Astronomical Observatories, Chinese Academy of Sciences. We acknowledge the support of the staff of Qingdao Aishan Observatory, Nanshan Observatory, Weihai Observatory, Skyline Observatory, Muztaga Observatory, Lijiang Gemini Observatory and Xingming Observatory. Some of the observations in this paper made use of the High-Resolution Imaging instrument 'Alopeke and were obtained under Gemini LLP Proposal Number: GN/S-2021A-LP-105. 'Alopeke was funded by the NASA Exoplanet Exploration Program and built at the NASA Ames Research Center by Steve B. Howell, Nic Scott, Elliott P. Horch, and Emmett Quigley. Alopeke was mounted on the Gemini North telescope of the international Gemini Observatory, a program of NSF's OIR Lab, which is managed by the Association of Universities for Research in cnomy (AURA) under a cooperative agreement with the National Science Foundation. on behalf of the Gemini partnership: the National Science Foundation (United States), National Research Council (Canada), Agencia Nacional de Investigaci\'on y Desarrollo (Chile), Ministerio de Ciencia, Tecnolog\'ia e Innovaci\'on (Argentina), Minist\'erio da Ci\^encia, Tecnologia, Inov\c{c}\~oes e Comunica\c{c}\~oes (Brazil), and Korea Astronomy and Space Science Institute (Republic of Korea). 

This work made use of Astropy\footnote{http://www.astropy.org}: a community-developed core Python package and an ecosystem of tools and resources for astronomy \citep{2013A&A...558A..33A,2018AJ....156..123A,2022ApJ...935..167A}. This work made use of Matplotlib\citep{2007CSE.....9...90H}, Numpy\citep{2011CSE....13b..22V}, Pyephem\citep{2011ascl.soft12014R} and Pytransit\citep{2015MNRAS.450.3233P}. This work has made use of data from the European Space Agency (ESA) mission {\it Gaia} (\url{https://www.cosmos.esa.int/gaia}), processed by the {\it Gaia} Data Processing and Analysis Consortium (DPAC,
\url{https://www.cosmos.esa.int/web/gaia/dpac/consortium}). Funding for the DPAC has been provided by national institutions, in particular the institutions participating in the {\it Gaia} Multilateral Agreement.

This paper makes use of EXOFAST\citep{2013PASP..125...83E} as provided by the NASA Exoplanet Archive, which is operated by the California Institute of Technology, under contract with the National Aeronautics and Space Administration under the Exoplanet Exploration Program.

\end{acknowledgements}

\bibliographystyle{raa}
\bibliography{bibtex}

\begin{thebibliography}{75}
\providecommand\natexlab[1]{#1}
\providecommand\JournalTitle[1]{#1}

\bibitem[{Akeson} {et~al.}(2013)]{2013PASP..125..989A}
{Akeson}, R.~L., {Chen}, X., {Ciardi}, D., {et~al.} 2013, \pasp, 125, 989

\bibitem[{Anders} {et~al.}(2019)]{2019A&A...628A..94A}
{Anders}, F., {Khalatyan}, A., {Chiappini}, C., {et~al.} 2019, \aap, 628, A94

\bibitem[{Anders} {et~al.}(2022)]{2022A&A...658A..91A}
{Anders}, F., {Khalatyan}, A., {Queiroz}, A.~B.~A., {et~al.} 2022, \aap, 658,
  A91

\bibitem[{Astropy Collaboration} {et~al.}(2013)]{2013A&A...558A..33A}
{Astropy Collaboration}, {Robitaille}, T.~P., {Tollerud}, E.~J., {et~al.} 2013,
  \aap, 558, A33

\bibitem[{Astropy Collaboration} {et~al.}(2018)]{2018AJ....156..123A}
{Astropy Collaboration}, {Price-Whelan}, A.~M., {Sip{\H{o}}cz}, B.~M., {et~al.}
  2018, \aj, 156, 123

\bibitem[{Astropy Collaboration} {et~al.}(2022)]{2022ApJ...935..167A}
{Astropy Collaboration}, {Price-Whelan}, A.~M., {Lim}, P.~L., {et~al.} 2022,
  \apj, 935, 167

\bibitem[{Bai} {et~al.}(2018)]{2018RAA....18..107B}
{Bai}, C.-H., {Fu}, J.-N., {Li}, T.-R., {et~al.} 2018, Research in Astronomy
  and Astrophysics, 18, 107

\bibitem[{Beatty} {et~al.}(2012)]{2012ApJ...756L..39B}
{Beatty}, T.~G., {Pepper}, J., {Siverd}, R.~J., {et~al.} 2012, \apjl, 756, L39

\bibitem[{Bertin} \& {Arnouts}(1996)]{1996A&AS..117..393B}
{Bertin}, E., \& {Arnouts}, S. 1996, \aaps, 117, 393

\bibitem[{Borucki} {et~al.}(2010)]{2010Sci...327..977B}
{Borucki}, W.~J., {Koch}, D., {Basri}, G., {et~al.} 2010, Science, 327, 977

\bibitem[{Buchhave} {et~al.}(2010)]{2010ApJ...720.1118B}
{Buchhave}, L.~A., {Bakos}, G.~{\'A}., {Hartman}, J.~D., {et~al.} 2010, \apj,
  720, 1118

\bibitem[{Buchhave} {et~al.}(2012)]{2012Natur.486..375B}
{Buchhave}, L.~A., {Latham}, D.~W., {Johansen}, A., {et~al.} 2012, \nat, 486,
  375

\bibitem[{Buchhave} {et~al.}(2014)]{2014Natur.509..593B}
{Buchhave}, L.~A., {Bizzarro}, M., {Latham}, D.~W., {et~al.} 2014, \nat, 509,
  593

\bibitem[{Charbonneau} {et~al.}(2000)]{2000ApJ...529L..45C}
{Charbonneau}, D., {Brown}, T.~M., {Latham}, D.~W., \& {Mayor}, M. 2000, \apjl,
  529, L45

\bibitem[{Collins} {et~al.}(2017)]{2017AJ....153...77C}
{Collins}, K.~A., {Kielkopf}, J.~F., {Stassun}, K.~G., \& {Hessman}, F.~V.
  2017, \aj, 153, 77

\bibitem[{Cooke} {et~al.}(2018)]{2018A&A...619A.175C}
{Cooke}, B.~F., {Pollacco}, D., {West}, R., {McCormac}, J., \& {Wheatley},
  P.~J. 2018, \aap, 619, A175

\bibitem[{Csizmadia} {et~al.}(2013)]{2013A&A...549A...9C}
{Csizmadia}, S., {Pasternacki}, T., {Dreyer}, C., {et~al.} 2013, \aap, 549, A9

\bibitem[{Deleuil} \& {Fridlund}(2018)]{2018haex.bookE..79D}
{Deleuil}, M., \& {Fridlund}, M. 2018, in Handbook of Exoplanets, ed. H.~J.
  {Deeg} \& J.~A. {Belmonte}, 79

\bibitem[{Demory} \& {Seager}(2011)]{2011ApJS..197...12D}
{Demory}, B.-O., \& {Seager}, S. 2011, \apjs, 197, 12

\bibitem[{Djupvik} \& {Andersen}(2010)]{2010ASSP...14..211D}
{Djupvik}, A.~A., \& {Andersen}, J. 2010, in Astrophysics and Space Science
  Proceedings, Vol.~14, Highlights of Spanish Astrophysics V, 211

\bibitem[{Dressing} \& {Charbonneau}(2015)]{2015ApJ...807...45D}
{Dressing}, C.~D., \& {Charbonneau}, D. 2015, \apj, 807, 45

\bibitem[{Droege} {et~al.}(2006)]{2006PASP..118.1666D}
{Droege}, T.~F., {Richmond}, M.~W., {Sallman}, M.~P., \& {Creager}, R.~P. 2006,
  \pasp, 118, 1666

\bibitem[{Eastman}(2017)]{2017ascl.soft10003E}
{Eastman}, J. 2017, {EXOFASTv2: Generalized publication-quality exoplanet
  modeling code}, Astrophysics Source Code Library, record ascl:1710.003

\bibitem[{Eastman} {et~al.}(2013)]{2013PASP..125...83E}
{Eastman}, J., {Gaudi}, B.~S., \& {Agol}, E. 2013, \pasp, 125, 83

\bibitem[{Espinoza} {et~al.}(2017)]{2017MNRAS.471.4374E}
{Espinoza}, N., {Rabus}, M., {Brahm}, R., {et~al.} 2017, \mnras, 471, 4374

\bibitem[{Fan} {et~al.}(2016)]{2016PASP..128k5005F}
{Fan}, Z., {Wang}, H., {Jiang}, X., {et~al.} 2016, \pasp, 128, 115005

\bibitem[{Foreman-Mackey} {et~al.}(2013)]{2013PASP..125..306F}
{Foreman-Mackey}, D., {Hogg}, D.~W., {Lang}, D., \& {Goodman}, J. 2013, \pasp,
  125, 306

\bibitem[{Fressin} {et~al.}(2013)]{2013ApJ...766...81F}
{Fressin}, F., {Torres}, G., {Charbonneau}, D., {et~al.} 2013, \apj, 766, 81

\bibitem[{F{\"u}r{\'e}sz}(2008)]{Frsz2008DESIGNAA}
{F{\"u}r{\'e}sz}, G. 2008, {DESIGN AND APPLICATION OF HIGH RESOLUTION AND
  MULTIOBJECT SPECTROGRAPHS: DYNAMICAL STUDIES OF OPEN CLUSTERS}, Ph.d thesis,
  UNIVERSITY OF SZEGED

\bibitem[{Gaia Collaboration}(2018)]{2018yCat.1345....0G}
{Gaia Collaboration}. 2018, VizieR Online Data Catalog, I/345

\bibitem[{Gaia Collaboration}(2022)]{2022yCat.1357....0G}
{Gaia Collaboration}. 2022, VizieR Online Data Catalog, I/357

\bibitem[{Gaia Collaboration} {et~al.}(2023)]{2023A&A...674A...1G}
{Gaia Collaboration}, {Vallenari}, A., {Brown}, A.~G.~A., {et~al.} 2023, \aap,
  674, A1

\bibitem[{Gao} {et~al.}(2016)]{2016PASP..128l5002G}
{Gao}, D.-Y., {Ji}, H.-X., {Cao}, C., {et~al.} 2016, \pasp, 128, 125002

\bibitem[{Grieves} {et~al.}(2021)]{2021A&A...652A.127G}
{Grieves}, N., {Bouchy}, F., {Lendl}, M., {et~al.} 2021, \aap, 652, A127

\bibitem[{Hopp} {et~al.}(2014)]{2014SPIE.9145E..2DH}
{Hopp}, U., {Bender}, R., {Grupp}, F., {et~al.} 2014, in Society of
  Photo-Optical Instrumentation Engineers (SPIE) Conference Series, Vol. 9145,
  Ground-based and Airborne Telescopes V, ed. L.~M. {Stepp}, R.~{Gilmozzi}, \&
  H.~J. {Hall}, 91452D

\bibitem[{Horne}(2003)]{2003ASPC..294..361H}
{Horne}, K. 2003, in Astronomical Society of the Pacific Conference Series,
  Vol. 294, Scientific Frontiers in Research on Extrasolar Planets, ed.
  D.~{Deming} \& S.~{Seager}, 361

\bibitem[{Howell}(2006)]{2006hca..book.....H}
{Howell}, S.~B. 2006, {Handbook of CCD Astronomy}, Vol.~5

\bibitem[{Howell} {et~al.}(2011)]{2011AJ....142...19H}
{Howell}, S.~B., {Everett}, M.~E., {Sherry}, W., {Horch}, E., \& {Ciardi},
  D.~R. 2011, \aj, 142, 19

\bibitem[{Howell} {et~al.}(2021)]{2021AJ....161..164H}
{Howell}, S.~B., {Matson}, R.~A., {Ciardi}, D.~R., {et~al.} 2021, \aj, 161, 164

\bibitem[{Howell} {et~al.}(2014)]{2014PASP..126..398H}
{Howell}, S.~B., {Sobeck}, C., {Haas}, M., {et~al.} 2014, \pasp, 126, 398

\bibitem[{Hu} {et~al.}(2014)]{2014RAA....14..719H}
{Hu}, S.-M., {Han}, S.-H., {Guo}, D.-F., \& {Du}, J.-J. 2014, Research in
  Astronomy and Astrophysics, 14, 719

\bibitem[{Huang} {et~al.}(2012)]{2012RAA....12.1585H}
{Huang}, F., {Li}, J.-Z., {Wang}, X.-F., {et~al.} 2012, Research in Astronomy
  and Astrophysics, 12, 1585

\bibitem[{Hunter}(2007)]{2007CSE.....9...90H}
{Hunter}, J.~D. 2007, Computing in Science and Engineering, 9, 90

\bibitem[{Jenkins} {et~al.}(2016)]{2016SPIE.9913E..3EJ}
{Jenkins}, J.~M., {Twicken}, J.~D., {McCauliff}, S., {et~al.} 2016, in Society
  of Photo-Optical Instrumentation Engineers (SPIE) Conference Series, Vol.
  9913, Software and Cyberinfrastructure for Astronomy IV, ed. G.~{Chiozzi} \&
  J.~C. {Guzman}, 99133E

\bibitem[{Kaltenegger} \& {Traub}(2009)]{2009ApJ...698..519K}
{Kaltenegger}, L., \& {Traub}, W.~A. 2009, \apj, 698, 519

\bibitem[{Kanodia} \& {Wright}(2018)]{2018ascl.soft08001K}
{Kanodia}, S., \& {Wright}, J.~T. 2018, {Barycorrpy: Barycentric velocity
  calculation and leap second management}, Astrophysics Source Code Library,
  record ascl:1808.001

\bibitem[{Kharchenko}(2001)]{2001KFNT...17..409K}
{Kharchenko}, N.~V. 2001, Kinematika i Fizika Nebesnykh Tel, 17, 409

\bibitem[{Kipping}(2010)]{2010MNRAS.408.1758K}
{Kipping}, D.~M. 2010, \mnras, 408, 1758

\bibitem[{Kurucz}(1992)]{1992IAUS..149..225K}
{Kurucz}, R.~L. 1992, in The Stellar Populations of Galaxies, ed. B.~{Barbuy}
  \& A.~{Renzini}, Vol. 149, 225

\bibitem[{Lester} {et~al.}(2021{\natexlab{a}})]{2021AAS...23734401L}
{Lester}, K., {Howell}, S., {Ciardi}, D., {et~al.} 2021{\natexlab{a}}, in
  American Astronomical Society Meeting Abstracts, Vol.~53, American
  Astronomical Society Meeting Abstracts, 344.01

\bibitem[{Lester} {et~al.}(2021{\natexlab{b}})]{2021AJ....162...75L}
{Lester}, K.~V., {Matson}, R.~A., {Howell}, S.~B., {et~al.} 2021{\natexlab{b}},
  \aj, 162, 75

\bibitem[{Mazeh} {et~al.}(2016)]{2016A&A...589A..75M}
{Mazeh}, T., {Holczer}, T., \& {Faigler}, S. 2016, \aap, 589, A75

\bibitem[{Miller} \& {Fortney}(2011)]{2011ApJ...736L..29M}
{Miller}, N., \& {Fortney}, J.~J. 2011, \apjl, 736, L29

\bibitem[{Morris} {et~al.}(2020)]{2020ksci.rept....6M}
{Morris}, R.~L., {Twicken}, J.~D., {Smith}, J.~C., {et~al.} 2020, {Kepler Data
  Processing Handbook: Photometric Analysis}, Kepler Science Document
  KSCI-19081-003, id. 6. Edited by Jon M. Jenkins.

\bibitem[{Naef} {et~al.}(2001)]{2001A&A...375L..27N}
{Naef}, D., {Latham}, D.~W., {Mayor}, M., {et~al.} 2001, \aap, 375, L27

\bibitem[{Narita} {et~al.}(2019)]{2019JATIS...5a5001N}
{Narita}, N., {Fukui}, A., {Kusakabe}, N., {et~al.} 2019, Journal of
  Astronomical Telescopes, Instruments, and Systems, 5, 015001

\bibitem[{Owen} \& {Wu}(2013)]{2013ApJ...775..105O}
{Owen}, J.~E., \& {Wu}, Y. 2013, \apj, 775, 105

\bibitem[{Paegert} {et~al.}(2021)]{2021arXiv210804778P}
{Paegert}, M., {Stassun}, K.~G., {Collins}, K.~A., {et~al.} 2021, arXiv
  e-prints, arXiv:2108.04778

\bibitem[{Parviainen}(2015)]{2015MNRAS.450.3233P}
{Parviainen}, H. 2015, \mnras, 450, 3233

\bibitem[{Pont} {et~al.}(2009)]{2009A&A...502..695P}
{Pont}, F., {H{\'e}brard}, G., {Irwin}, J.~M., {et~al.} 2009, \aap, 502, 695

\bibitem[{Rasio} {et~al.}(1992)]{1992Natur.355..325R}
{Rasio}, F.~A., {Nicholson}, P.~D., {Shapiro}, S.~L., \& {Teukolsky}, S.~A.
  1992, \nat, 355, 325

\bibitem[{Steuer} {et~al.}(2021)]{2021SPIE11823E..1US}
{Steuer}, J., {Kellermann}, H., {Grupp}, F., {et~al.} 2021, in Society of
  Photo-Optical Instrumentation Engineers (SPIE) Conference Series, Vol. 11823,
  Techniques and Instrumentation for Detection of Exoplanets X, ed. S.~B.
  {Shaklan} \& G.~J. {Ruane}, 118231U

\bibitem[{Szentgyorgyi} {et~al.}(2005)]{2005AAS...20711010S}
{Szentgyorgyi}, A.~H., {Geary}, J.~G., {Latham}, D.~W., {et~al.} 2005, in
  American Astronomical Society Meeting Abstracts, Vol. 207, American
  Astronomical Society Meeting Abstracts, 110.10

\bibitem[{Telting} {et~al.}(2014)]{2014AN....335...41T}
{Telting}, J.~H., {Avila}, G., {Buchhave}, L., {et~al.} 2014, Astronomische
  Nachrichten, 335, 41

\bibitem[{Tingley}(2004)]{2004A&A...425.1125T}
{Tingley}, B. 2004, \aap, 425, 1125

\bibitem[{Tody}(1986)]{1986SPIE..627..733T}
{Tody}, D. 1986, in Society of Photo-Optical Instrumentation Engineers (SPIE)
  Conference Series, Vol. 627, Instrumentation in astronomy VI, ed. D.~L.
  {Crawford}, 733

\bibitem[{Trifonov}(2019)]{2019ascl.soft06004T}
{Trifonov}, T. 2019, {The Exo-Striker: Transit and radial velocity interactive
  fitting tool for orbital analysis and N-body simulations}, Astrophysics
  Source Code Library, record ascl:1906.004

\bibitem[{Tuomi} {et~al.}(2014)]{2014MNRAS.441.1545T}
{Tuomi}, M., {Jones}, H. R.~A., {Barnes}, J.~R., {Anglada-Escud{\'e}}, G., \&
  {Jenkins}, J.~S. 2014, \mnras, 441, 1545

\bibitem[{Tuomi} {et~al.}(2019)]{2019arXiv190604644T}
{Tuomi}, M., {Jones}, H.~R.~A., {Butler}, R.~P., {et~al.} 2019, arXiv e-prints,
  arXiv:1906.04644

\bibitem[{Twicken} {et~al.}(2010)]{2010SPIE.7740E..23T}
{Twicken}, J.~D., {Clarke}, B.~D., {Bryson}, S.~T., {et~al.} 2010, in Society
  of Photo-Optical Instrumentation Engineers (SPIE) Conference Series, Vol.
  7740, Software and Cyberinfrastructure for Astronomy, ed. N.~M. {Radziwill}
  \& A.~{Bridger}, 774023

\bibitem[{van der Walt} {et~al.}(2011)]{2011CSE....13b..22V}
{van der Walt}, S., {Colbert}, S.~C., \& {Varoquaux}, G. 2011, Computing in
  Science and Engineering, 13, 22

\bibitem[{Winn}(2010)]{2010exop.book...55W}
{Winn}, J.~N. 2010, in Exoplanets, ed. S.~{Seager}, 55

\bibitem[{Zhang} {et~al.}(2015)]{2015PASP..127.1292Z}
{Zhang}, J.-C., {Ge}, L., {Lu}, X.-M., {et~al.} 2015, \pasp, 127, 1292

\bibitem[{Zhao} {et~al.}(2019)]{2019MNRAS.482.1406Z}
{Zhao}, F., {Zhao}, G., {Liu}, Y., {et~al.} 2019, \mnras, 482, 1406

\bibitem[{Zheng}(2023)]{2023ART...01...83}
{Zheng}, J., J. L.~Q. 2023, Astronomical Research \& Technology, 20(01), 83

\end{thebibliography}

\label{lastpage}

\end{document}